\documentclass[
reprint,
amsmath,amssymb,
aps, groupedaddress,superscriptaddress
]{revtex4-1}

\usepackage{graphicx}  
\usepackage{bm}        
\usepackage{verbatim}

\renewcommand{\vec}[1]{\mathbf{#1}}
\newcommand{\bk}{\vec{k}}
\newcommand{\bq}{\vec{q}}
\newcommand{\br}{\vec{r}}
\newcommand{\bJ}{\vec{J}}
\newcommand{\qorder}{Q}
\newcommand{\bqorder}{\vec{Q}}
\newcommand{\brecip}{\vec{G}}
\newcommand{\recip}{G}
\newcommand{\epk}{\epsilon_{\bk}}
\newcommand{\ephatk}{\epsilon_{\hat\bk}}
\newcommand{\epik}{\epsilon_{i\bk}}
\newcommand{\epjk}{\epsilon_{j\bk}}
\newcommand{\epq}{\epsilon_{\qorder}}
\newcommand{\nbar}{\bar{n}}
\newcommand{\ncell}{n_{\text{cell}}}
\newcommand{\thetabar}{\bar{\theta}}
\newcommand{\dn}{\delta n}
\newcommand{\dtheta}{\delta\theta}
\newcommand{\Vtilde}{\tilde{V}}
\newcommand{\Egap}{E_{\text{g}}}
\newcommand{\Jperp}{J_\perp}

\begin{document}

\title{Intertwined Superfluidity and Density Wave Order in a
  $p$-Orbital Bose Condensate}
\author{Simon Lieu}
\affiliation{
   Blackett Laboratory, Imperial College London, London SW7 2AZ, United Kingdom
   }
\author{Andrew F.~Ho}
\affiliation{
   Department of Physics, Royal Holloway, University of London, Egham, Surrey TW20 0EX, United Kingdom
   }
   
   \author{Derek K.~K.~Lee}
\affiliation{
   Blackett Laboratory, Imperial College London, London SW7 2AZ, United Kingdom
   }

   \author{Piers Coleman}
\affiliation{
   Center for Materials Theory, Rutgers University, Piscataway, New Jersey, 08854, USA
   }   
   \affiliation{
   Department of Physics, Royal Holloway, University of London, Egham, Surrey TW20 0EX, United Kingdom
   }   
   
\date{\today}
\begin{abstract}
We study a continuum model of the weakly interacting Bose gas in the presence of an external field with minima forming a triangular lattice. The second lowest band of the single-particle spectrum ($p$-band) has three minima at non-zero momenta. We consider a metastable Bose condensate at these momenta and find that, in the presence of interactions that vary slowly over the lattice spacing, the order parameter space is isomorphic to $S^{5}$. We show that the enlarged symmetry leads to the loss of topologically stable vortices, as well as two extra gapless modes with quadratic dispersion. The former feature implies that this non-Abelian condensate is a ``failed superfluid'' that does not undergo a Berezinskii-Kosterlitz-Thouless (BKT) transition.  Order-by-disorder splitting appears suppressed, implying that signatures of the $S^5$ manifold ought to be observable at low temperatures.
\end{abstract}

\maketitle

\section{Introduction}

The search for novel quantum states of matter continues to be a central theme in condensed matter physics. The conventional Bose superfluid spontaneously breaks a continuous U(1) symmetry in the global phase of the condensate wavefunction, giving rise to quantized vortices as a signature of the system. Motivated by recent experiments on two completely different platforms, we investigate the possibility of intertwining superfluid order with density wave order as a result of additional degeneracies in the spatial structure of a Bose-Einstein condensate (BEC). In this paper, we present  a simple theory of an intertwined state that has a superfluid stiffness, but has a non-Abelian order parameter manifold that lacks the topological protection to support superflow at non-zero temperatures. 
  
Our first experimental motivation comes from torsional oscillator experiments on $^4$He bilayers on graphite suggesting that a 2D superfluid in a periodic potential may exhibit an unconventional quantum phase \cite{Nyeki2017}, notably characterized by the lack of a BKT transition \cite{Berezinskii1971,*Kosterlitz1973} and a linear temperature dependence of the normal fraction (\emph{c.f.}~cubic behavior in the conventional superfluid). The system is close to an incommensurate solid phase. An order parameter was postulated \cite{Nyeki2017} with intertwined superfluidity and crystallinity such that the global phase and translational degrees of freedom are no longer independent.  This order parameter exists on an enlarged symmetry manifold. The BKT transition would be eliminated since global phase vortices would not be topologically protected, while unconventional Goldstone modes could be the source of an enhanced normal fraction.

Secondly, there has also been a series of remarkable experiments \cite{Landig2016,*Leonard2016} that succeeded in creating a spatially modulated atomic BEC in an optical lattice that arose from the spontaneous occupation of photon cavity modes in two intersecting cavities. A new U(1) Goldstone mode was observed associated with a degenerate set of density wave patterns. This again raises the tantalizing possibility of intertwined density wave order and superfluidity.

This paper is organized as follows. In section
\ref{sec:model}, we introduce a model bosonic Hamiltonian with
long-range interactions. In section \ref{sec:singleparticle}, we
examine its band structure in the non-interacting limit and focus on
the second lowest band where bosons can condense into three degenerate
single-particle states at zero temperature, which can be described as
a coherent state. In section \ref{sec:nobkt}, we argue that the
coherent state exists on a $S^5$ manifold, in contrast to the
conventional U(1) manifold where condensation only occurs in one
single-particle state. Using general arguments, we argue that we do
not expect the system to be a superfluid at non-zero temperatures in
the absence of topological protection of the vortices on an $S^5$
manifold. We then proceed to support this conjecture by calculating
the excitation spectra of the system at zero temperature (section
\ref{sec:spectra}) and then using this to verify that both the
condensate depletion and normal fluid density diverge in the
thermodynamic limit at non-zero temperatures. In sections
\ref{sec:orderbydisorder} and \ref{sec:bktsuppress}, we discuss the
validity of our conclusions when the $S^5$ symmetry is weakly broken. Finally, we
discuss the implications of our work in section \ref{sec:conclude}.

\section{The model}
\label{sec:model}

In this paper, we study a simple bosonic Hamiltonian to understand such an intertwined quantum state. We will see that two ingredients are required. We need degenerate single-particle states with different spatial structures to form an enlarged order parameter space. To preserve this degeneracy, the interaction between particles has to be smooth and long-ranged. We consider bosons in 2D with mass $m$ in an external potential: $U(\br) = 2U \sum_{j=1}^3 \sin^2(\brecip_j\cdot\vec{r}/2)$ where $\brecip_{j}=G (\cos\theta_j, \sin \theta_j)$ with $\theta_j=\pi(j-1)/3$ $(j=1,...,6)$ are the reciprocal lattice vectors. We take $U>0$ which gives potential minima on a triangular lattice  with lattice constant $a=4\pi/\sqrt{3}G$. The Hamiltonian is
\begin{multline}
\label{eqn:Ham}
 H=\sum_{\bk}\left(\epk -\mu\right)b_{\bk}^{\dagger}b_{\bk}-\frac{U}{2}\sum_{j,\bk}
\left( b_{\bk+\brecip_j}^\dagger b_{\bk} +\text{h.c.} \right) \\
+\frac{1}{2L^2}\sum_{\bk,\bk',\bq}V_{\bq}b_{\bk-\bq}^{\dagger}b_{\bk'+\bq}^{\dagger}b_{\bk'}b_{\bk}
\end{multline}
where $\epk = \hbar^2 k^2/2m$, $\mu$ is the chemical potential, $V_{\bq}$ is the Fourier transform of the interaction potential, $L$ is the linear size of the system and $b_{\bk}$ is the bosonic annihilation operator.

For the degeneracy requirement, we exploit the $p$-band of this system which has degenerate minima in the first Brillouin zone (see below). We note that $p$-band BECs can be realized experimentally with ultracold atoms  \cite{Mueller2007,Wirth2011} with novel features due to symmetries from ``internal'' degrees of freedom, \emph{e.g.}~orbitals or spins \cite{Isacsson2005,Liu2006}. Our proposal is different in that we make use of degeneracies in spatial structure rather than internal symmetries.

We will now discuss the requirement on the form of the
interactions. Theoretical studies of ultracold atoms typically employ
a zero-range contact interaction to model $s$-wave scattering because
the range of the interaction is short compared to the wavelength of
the condensate. In this work, we are interested in a condensate with
spatial modulation commensurate with a wavevector of magnitude
$\recip/2$. We will study interactions $V_\bq$ that are smooth over a
length scale $R$ that is long compared to $1/\recip$, such that there
are momentum transfers of the order of $\recip$ are suppressed. Smooth
interactions can be realized in dipolar atoms/molecules
\cite{Fischer2006} by tuning a Feshbach resonance so that the contact
repulsion is cancelled by the dipolar attraction at short range (see
Appendix \ref{app:dipolar}). The cooling of dipolar atoms is
challenging but an Er BEC in an optical lattice has been achieved
recently \cite{Baier2016,Chomaz2018,Petter2018}. For
our calculations below, we use the simple mathematical form: $V_\bq =
V_0 \exp(-q^2R^2)$ with $\recip R\gg 1$ for the convenience of a
model with only a single parameter $GR$ to control the smoothness of
the interaction. We will focus on results that are insensitive to the
precise form of the interaction.

\section{Single-particle structure and the $S^5$ manifold}
\label{sec:singleparticle}

\begin{figure}[tb]
{\includegraphics[width= 0.3\columnwidth]{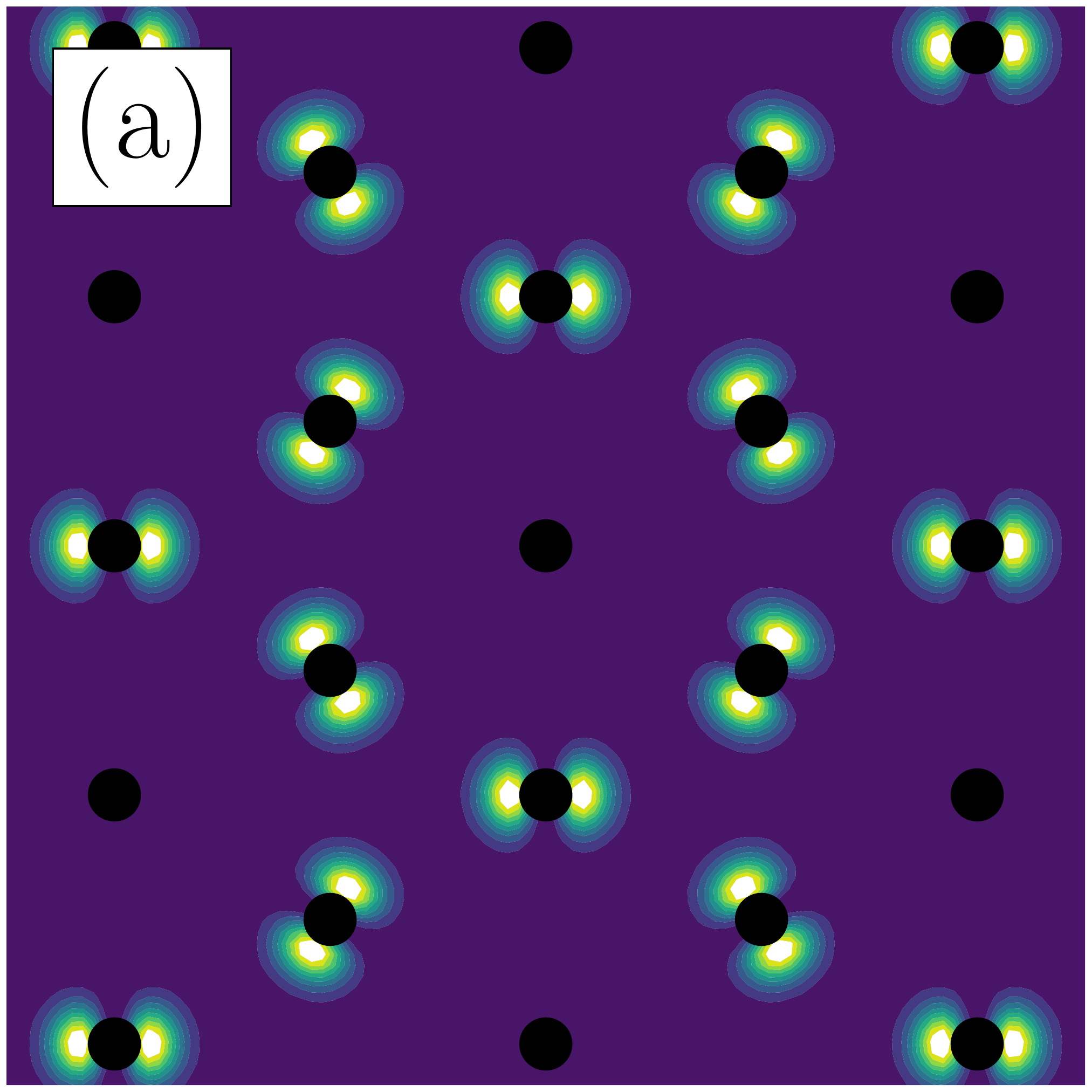}}
{\includegraphics[width= 0.3\columnwidth]{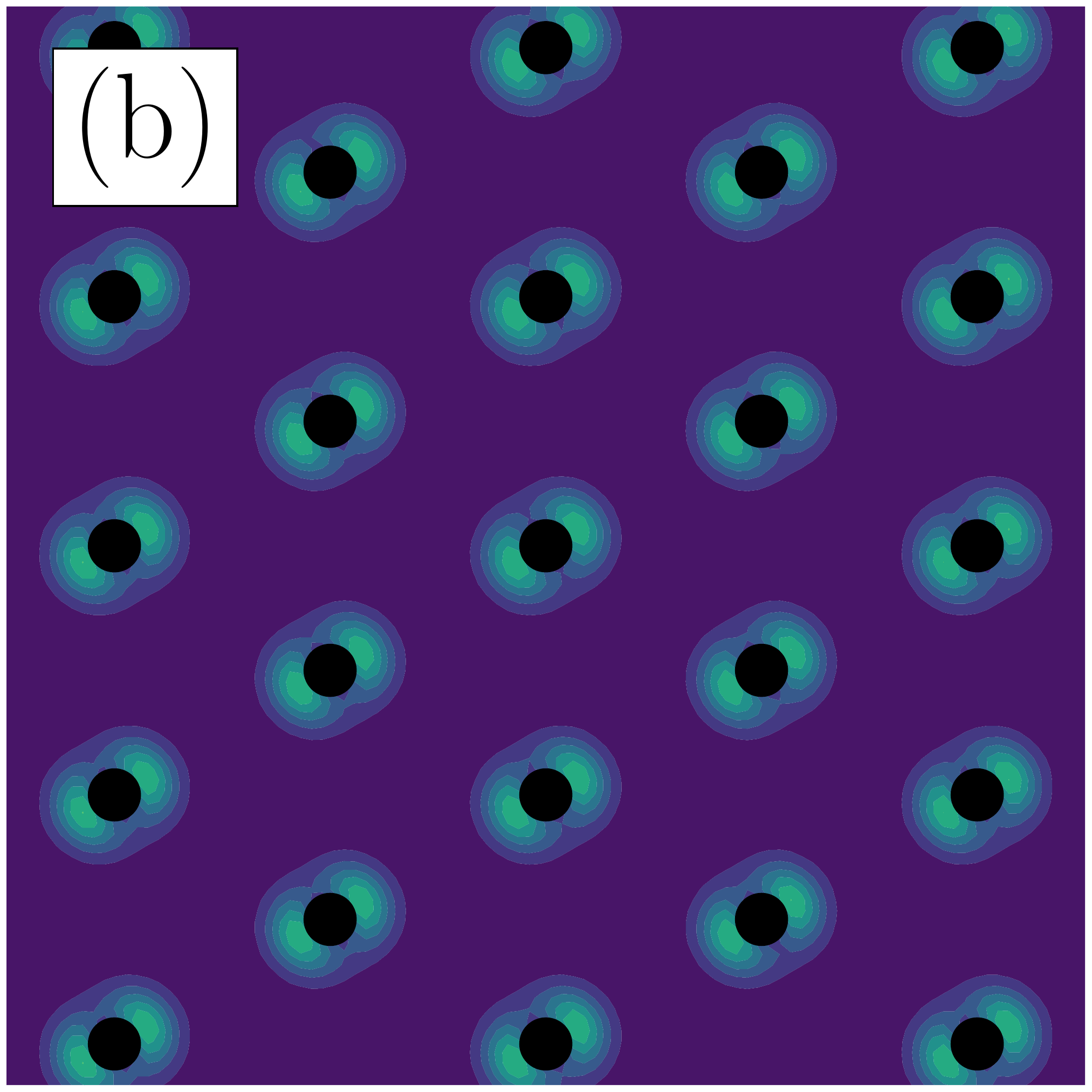}}
\includegraphics[width=0.33\columnwidth]{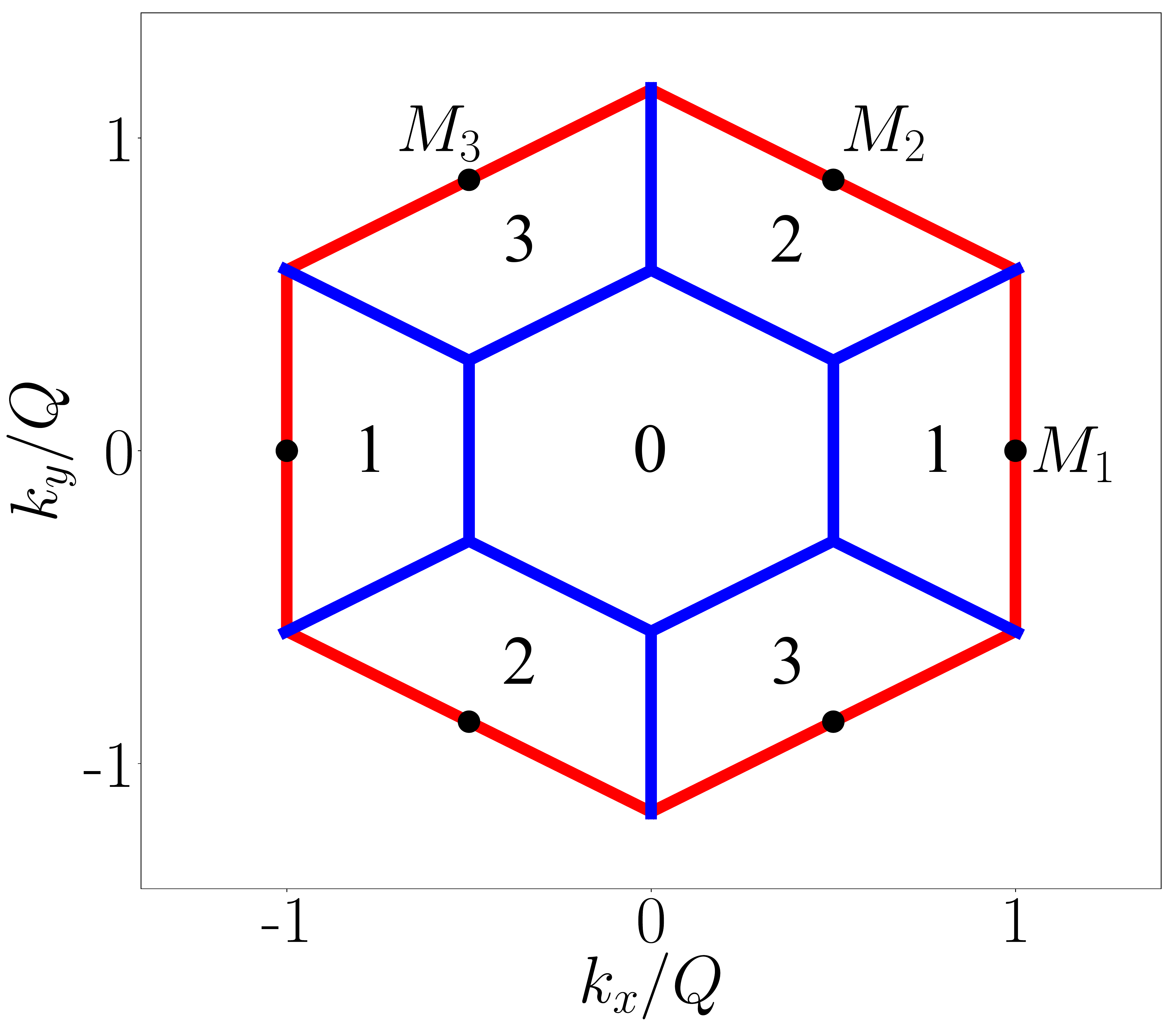}
\caption{(a) Density profile for $(c_1,c_2,c_3)=(1,1,1)/\sqrt{3}$ has  point-group symmetry of honeycomb lattice.  States with equivalent densities (up to  a translation) can be constructed by changing the sign   of one of the amplitudes, leading to a $\mathbb{Z}_4$  symmetry; (b) Density profile of $(c_1,c_2,c_3)=(1,i,0)/\sqrt{2}$ has a $C_{2v}$ point  group. Black dots: potential minima. (c) Brillouin zone of the external potential (red) with reciprocal lattice vectors $\brecip_{i}$. The $p$-band minima occur at three inequivalent $M$ points at $\bqorder_i=\brecip_i/2$. Upon condensation at the $M$ points, the spatial periodicity of the system doubles, and excitations have a reduced Brillouin zone (blue).
}\label{fig:density}
\end{figure}

Let us examine now the single-particle band structure ($V_{\bq}=0$) of
this system. The lowest band has the lowest energy at the $\Gamma$
point. We will focus on the next lowest band which corresponds in the
tight-binding limit to Bloch states formed by $p_x$ and $p_y$ orbitals
in each well of the external potential.  This $p$-band has three
degenerate local minima (due to rotational symmetry) with energy
$\epsilon_M$ and crystal momenta $\bqorder_i\equiv \brecip_i/2$
($i=1,2,3$) at the three $M$ points ($M_{1,2,3}$) of the first
Brillouin zone of the triangular lattice
(Fig.~\ref{fig:density}c). The annihilation operators for these Bloch
states, $B_{1,2,3}$, can be written as a superposition of plane-wave
states connected by reciprocal lattice vectors $\brecip =
p_1\brecip_1+p_2\brecip_2$ for integer $p_{1,2}$: $B_i =
\sum_{\brecip}d_{\brecip,i}b_{\bqorder_i+\brecip}$. From these three
states, we can construct a degenerate set of single-particle states:
\begin{equation}
  (c_1 B_1^\dagger + c_2 B_2^\dagger +c_3 B_3^\dagger)\left|\text{vac}
  \right\rangle,\,
  |c_1|^2+|c_2|^2+|c_3|^2=1,
\end{equation}
each of which can be represented as a point on the $S^5$ surface in
$\mathbb{R}^6$. Depending on the magnitude and phases of the $c_i$'s,
these degenerate states have quite different density profiles,
generically breaking completely the point-group symmetries of the
external potential. We will study the condensation of bosons into
these states. Two of the more symmetric cases are plotted in
Fig.~\ref{fig:density}.

\begin{figure}[tb]
\vspace{-\baselineskip}
{\includegraphics[width= 0.28\columnwidth]{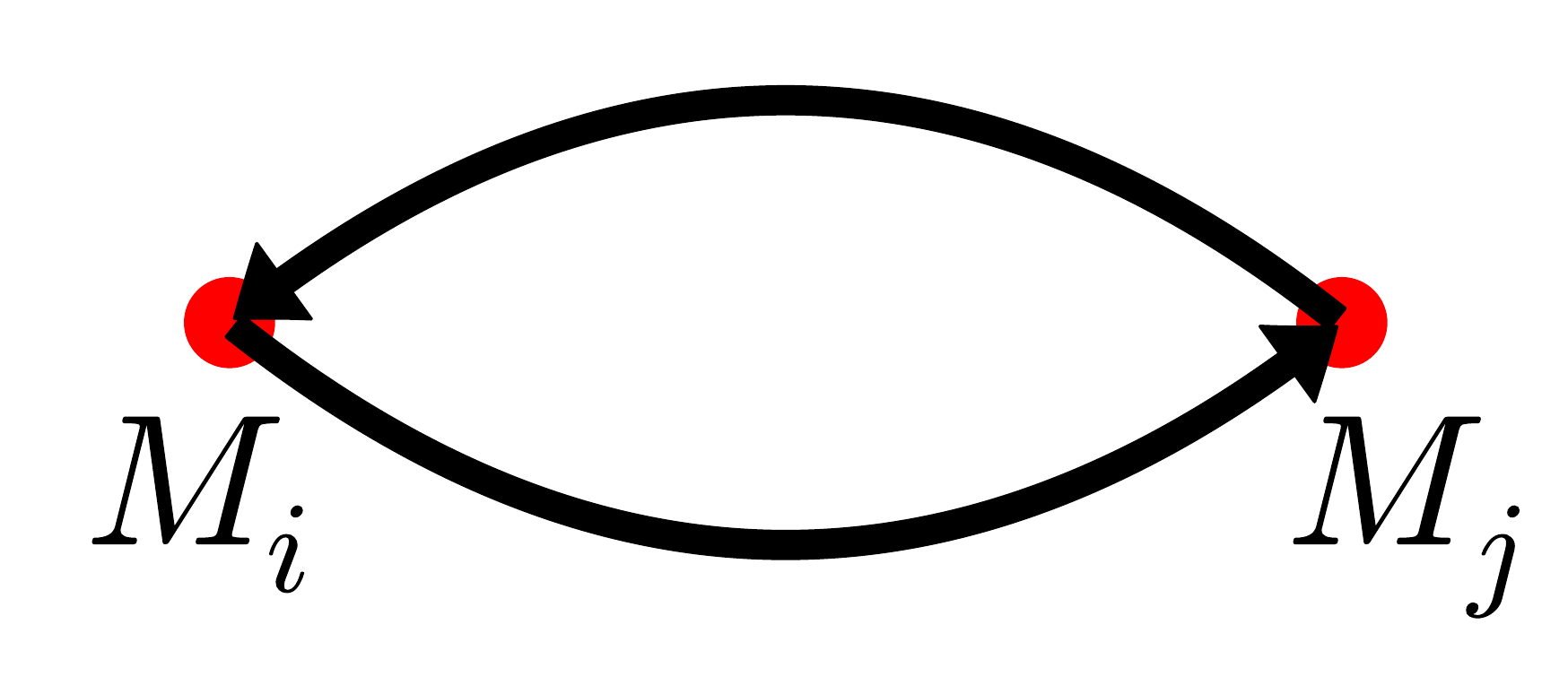}}\hspace{1em}
{\includegraphics[width= 0.28\columnwidth]{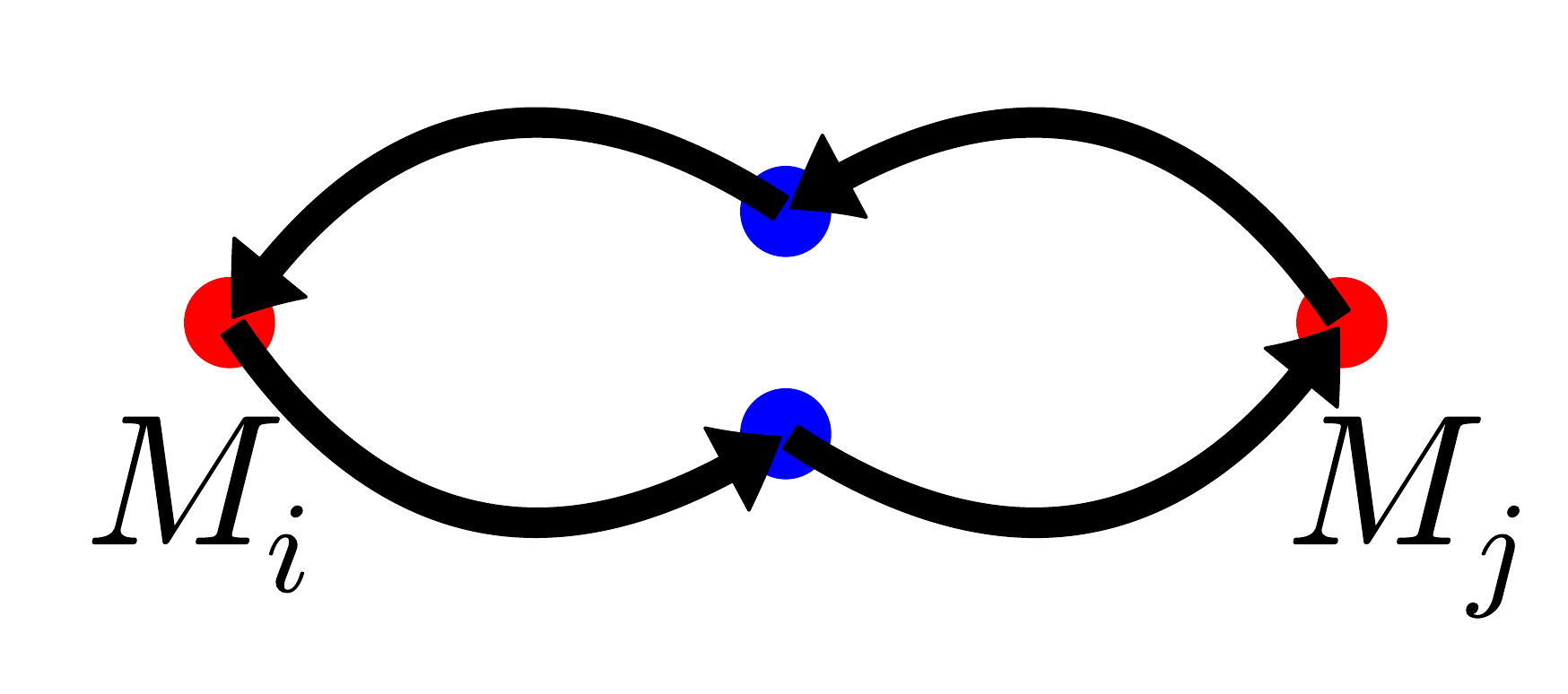}}
\caption{Direct  exchange (left) and superexchange (right) between
  condensed particles at momenta $M_i$ and $M_j$. Direct exchange is
  absent if $V_Q=0$.}\label{fig:exchange}
\end{figure}

We now turn to the effect of interactions.  One would expect
short-ranged interactions to select Bose condensation into a unique
member of this manifold \cite{Wu2009}. In this work, we aim to explore
the novel properties of a Bose-condensed system where the
single-particle degeneracy is preserved even in the presence of
interactions, by using an interaction that is smooth on the scale of
the lattice spacing $a$.  To be more precise, we consider the system
in a coherent state
of the form \cite{Nyeki2017,Henkel2010, Macri2013, Kunimi2012}:
\begin{equation}\label{eqn:manifold}
  \left|\Psi\right\rangle  = 
  e^{-N/2} \exp\left( \sqrt{N} \sum_{i=1}^3 c_{i}B_{{i}}^{\dagger}\right)\left|\text{vac}\right\rangle 
\end{equation}
where $N$ is the total number of particles in the system set by the
chemical potential $\mu$. By construction, this minimizes the
single-particle energy in the $p$-band.  The mean field energy for
this ansatz is of the form
$\langle\Psi|H|\Psi\rangle = N u_{\text{MF}}$:
\begin{equation}\label{eqn:hammf}
  \begin{split}
    u_{\text{MF}} &=
    \frac{\Vtilde_0\nbar}{2} 
    + \frac{{\Vtilde}_2\nbar}{4} \sum_{i=1}^3|c_{i}|^{4}\\
    &+ \frac{{\Vtilde}_1 \nbar}{4} \sum_{i=1}^3\left[2|c_{i}|^{2}|c_{i+1}|^{2}  +
      (c_{i}^{*2}c_{i+1}^{2} + \text{c.c.})\right]
  \end{split}
\end{equation}
where the addition in the $i$-index is modulo 3, $\nbar=N/L^2$ is the
average boson number. $\tilde{V}_1$ control intervalley scattering of
particles from one $M$ point to another (Fig.~\ref{fig:exchange} left)
and is the only term allowed by the conservation of crystal momentum.
In the case of $QR\gg 1$ (or $U\ll \epq$), the interaction strengths
can be approximated by $\Vtilde_0 \simeq V_0$,
$\Vtilde_1\simeq V_{\qorder}+V_{\sqrt{3}\qorder}$ and
$\Vtilde_2\simeq V_{2\qorder}$.

If there are strong interactions
commensurate with the wavevector $Q$, the system could become
insulating with a checkerboard pattern of site occupation if the
chemical potential is appropriately chosen to give a commensurate
particle density. This is analogous to
the Mott insulator for short-ranged interactions. This regime occurs if
$\tilde{V}_{1,2}$ becomes negative and larger than the bandwidth of the
$p$-band. We will not study this regime. Instead we will focus on
weak values of $\tilde{V}_{1,2}$. In this case, we expect the
system to condense into a coherent state and the key issue arising
from the interactions is then as follows. Generically, the interactions terms
($\Vtilde_{1,2}$) in $u_{\text{MF}}$ will be minimized for a
particular choice of the relative weights and phases for the
amplitudes $c_i$ in the coherent state \eqref{eqn:manifold}. This
leads to a reduction of the degenerate manifold from $S^{5}$ to the
conventional U(1) manifold. Indeed, for $\Vtilde_{1}\ge \Vtilde_2>0$,
Liu and Wu \cite{Liu2006,Wu2009} showed that the system favors
$c_1= \pm ic_2$, $c_3=0$ (and permutations) which has a U(1) symmetry
in the global phase as well as discrete symmetries: $\mathbb{Z}_2$ for
time reversal and $\mathbb{Z}_3$ for the choice of the empty state
(Fig.~\ref{fig:density}b).

In the spirit of preserving the degeneracy of the $S^{5}$ manifold
\eqref{eqn:manifold}, we specialize to a spatially smooth interaction
such that $V_{q\geq \qorder}=0$, {\it i.e.}
$\Vtilde_{1}=\Vtilde_2=0$.
The intervalley processes are absent and the
degeneracy on the $S^5$ manifold is not lifted by interactions (at the
mean field level).  The mean field energy also does not depend on the
relative phases of the amplitudes $c_i$ due to separate number
conservation at each $M$ point at this level.
In other words, ansatz \eqref{eqn:manifold}
minimizes both the single-particle and interaction terms in
$u_{\text{MF}}$ for any choice of $c_i$ as long as
$|c_1|^2+|c_2|^2+|c_3|^2=1$.

Henceforth, we will focus on this degenerate scenario where the
coherent state manifold \eqref{eqn:manifold} represents states of
minimal energy in the $p$-band. We should point
out that a similar
situation applies to the spontaneous optical lattice experiments
\cite{Landig2016,*Leonard2016} where the additional U(1) symmetry is
only approximate in the presence of $s$-wave scattering.

We discuss now the emergent SU(3) symmetry arising from the degeneracy
of the $S^5$ manifold (\ref{eqn:manifold}). The state
$(c_1 B_1^\dagger + c_2 B_2^\dagger +c_3 B_3^\dagger) \left|\text{vac}
\right>$ is a Schwinger boson formulation of the fundamental
representation of SU(3). The unitary transformation that connects any
two states in the manifold can be written as an SU(3) transformation
with the generators given by
$\Lambda^{a} = \sum_{i,j=1,2,3} B_i^\dagger \lambda_{i,j}^a B_j/2$,
where $\lambda^a$ are the eight Gell-Mann matrices as defined in
\cite{georgi1999lie}. This manifold may be parameterized via an SU(3)
rotation on the highest-weight state using the generators according
to:
\begin{equation}
\left|\Psi\right\rangle \propto \exp\left[ i \sum_{a=4}^8 \theta_a \Lambda^a  \right]  \exp\left[\sqrt{N}B_{3}^{\dagger}\right]\left|\text{vac}\right\rangle 
\end{equation}
Note that rotations involving the generators $\Lambda^{1,2,3}$ on the highest-weight state leaves it invariant and therefore the entire manifold may be parameterized parsimoniously using only five generators: $\Lambda^{4,\ldots,8}$, \emph{i.e.}~rotations involving the generators which form an SU(2) subgroup do not alter the highest-weight state. The manifold is isomorphic to SU(3)/SU(2) $\sim S^5$ \cite{Nemoto2002}.

\section{Lack of BKT transition on the $S^5$ manifold}
\label{sec:nobkt}

A non-Abelian symmetry manifold is interesting in the context of
superfluid BECs in two dimensions. To understand this, let us first review the
conventional theory.

Conventional superfluids with
condensation into one single-particle state has a complex scalar as
an order parameter. The global gauge symmetry of this order parameter
means that the coherent states span a U(1) manifold.
In two dimensions, the Mermin-Wagner-Hohenberg theorem
\cite{Mermin1966,*Hohenberg1967} forbids
spontaneous breaking of the U(1) symmetry, \emph{i.e.} no 
condensate with infinitely long ranged phase coherence, at any
non-zero temperatures for a system in the thermodynamic limit. However, the loss of
condensation does not necessarily lead to the loss of superfluidity.
The destruction of superfluidity in the
U(1) superfluid requires freely propagating vortices inducing phase
slips across the system. Berezinskii,
Kosterlitz and Thouless \cite{Berezinskii1971,*Kosterlitz1973} showed
that vortices are bound in vortex-antivortex pairs below a non-zero critical
temperature $T_{\rm BKT}$. The motion of these pairs do not cause
phase slips. Thus, the system remains in a superfluid phase up to
$T_{\rm BKT}$. Above this critical temperature, the vortices and
antivortex unbind and the system loses its superfluid response.

The BKT theory relies on the topological protection of
the vortices in the form of quantized circulation.
Since the BKT superfluid is the only known superfluid without
condensation at non-zero temperatures, 
this leads us to ask whether our non-Abelian condensate possesses
topologically protected defects which can protect it from phase slip events.
The first homotopy group
for our $S^5$ manifold is trivial: $\pi_{1}(S^{5})=0$, meaning that
any closed loop in the manifold of $S^{5}$ may be continuously shrunk
to a point. This means that phase vortices are not topologically
stable.  Explicitly, we can destroy the phase of the amplitude at
$M_i$ by a trajectory on the $S^5$ manifold that takes the coherent
state through a region where $c_i=0$. In addition, $\pi_{2}(S^{5})$ is
also trivial and so there are no topologically stable defects for our
coherent states in two dimensions.

In the absence of topological defects, we expect that our
system is not protected from thermally induced phase slips and so is
\emph{not} a superfluid in the thermodynamic limit. Thus, the 2D
$S^5$-degenerate condensate we study here is a rare example of an
interacting Bose system without superfluidity in the thermodynamic
limit at any non-zero temperature. In this sense, a condensate with
non-Abelian symmetry generators may be viewed as a ``failed
superfluid" in two dimensions, the failure being the lack of
topological protection when the order parameter exists on an $S^5$ manifold.

We should stress that the loss of condensate in two
dimensions only applies to the thermodynamic limit. Bose condensation
is possible in a finite system whose size is smaller than the
correlation length of the order parameter. We will see explicitly how a system
loses condensation and superfluidity as we increase the system size or
raise the temperature in section \ref{sec:scaling}. In order to
perform that stability analysis, we need to understand the low-energy
excitations of the system at zero temperature. This is studied in the
next section.

\section{Excitation spectra}
\label{sec:spectra}

In section \ref{sec:singleparticle}, we have an ansatz for the $S^5$
condensate at zero temperature. In this section, we will compute the
excitation spectrum at zero temperature. This is needed for us to
examine the stability of the condensate as we raise the temperature
from absolute zero in section \ref{sec:scaling}. This stability calculation will support our claims in section \ref{sec:nobkt}.

We outline two methods to study the excitation spectrum. The first method applies the Bogoliubov approximation to the full Hamiltonian and obtains numerical results for the excitations for multiple bands in the whole Brillouin zone. The second method describes an effective theory for long-wavelength fluctuations in the $p$-band of the system. This is useful in understanding the Goldstone modes of the system and the stability analysis in section \ref{sec:scaling}.


\subsection{Numerical Bogoliubov Calculation}
\label{subsec:numerical}

Consider first non-interacting bosons in the triangular potential $U(\br)$ [see (1)] consisting of Fourier components at the reciprocal lattice vectors $\pm \brecip_{1,2,3}$. For our numerical work, we work in the plane-wave (Fourier) basis. The eigenstates are Bloch states.  A Bloch state in the band $\gamma$ is created by the creation operator
\begin{equation}\label{eq:blochstate}
  (B^{(\gamma)}_\bq)^\dagger = \sum_{\brecip}d^{(\gamma)}_{\brecip,\bq} b^\dagger_{\brecip+\bq}\,.
\end{equation}
where $\brecip = p_1 \brecip_1 + p_2 \brecip_2$ with integer $p_{1,2}$ are the reciprocal lattice vectors, $b^\dagger_{\vec{p}}$ creates a plane-wave state at wavevector $\vec{p}$, and $\bq$ is restricted to the first Brillouin zone of the non-interacting problem (red hexagon in Fig.~\ref{fig:density}c). In other words, it is a superposition of plane waves with wavevectors separated by reciprocal lattice vectors $\brecip$. There are two $p$-bands, the lower one of which has energy minima at the $M_{1,2,3}$ points of Brillouin zone corresponding to $\bq = \bqorder_{1,2,3} = \brecip_{1,2,3}/2$. We will from now on refer to this lower band as ``the $p$-band''.

We are concerned with the excitation spectrum after the particles have Bose-condensed into  the Bloch states at the three $M$ points. The spatial modulation of the condensate has Fourier components at integer multiples of $\bqorder_{1,2,3}$. This means that the Brillouin zone for the excitations is halved in each direction (Fig.~\ref{fig:density}c). So, it is more convenient to label states in four reduced Brillouin zones ($m=0,1,2,3$, shown in Fig.~\ref{fig:density}c). Let us also divide all the plane-wave states into reduced Brillouin zones centered at $\bqorder_m = p_1\bqorder_1+p_2\bqorder_2$ for some integers $p_{1,2}$. (The numerical calculation cuts off the basis at $m=m_c$.)  Let us denote the creation operator for a free particle with wavevector $\bqorder_m+\bk$ and energy $\hbar^2(\bqorder_m+\bk)^2/2m$ as $b^\dagger_{m,\bk}\equiv b^\dagger_{\bqorder_m+\bk}$ where $\bk$ is restricted to the first Brillouin zone (central $m=0$ blue hexagon in Fig.~\ref{fig:density}c) of the reduced Brillouin zones. In this work, we will focus on condensation into the $p$-band which has energy minima at the $M_{1,2,3}$ points (Fig.~\ref{fig:density}c). The condensate creation operator can be written as:
\begin{equation}
  c_1 B^\dagger_{1,\bk=0}+c_2 B^\dagger_{2,\bk=0}+c_3
  B^\dagger_{3,\bk=0} = \sum_m\alpha_{m}  b^\dagger_{m,\bk=0}
\end{equation}
where $B^\dagger_{j,\bk}$ creates a Bloch state with crystal momentum $\bk$ near the $M_j$ point. (For each $j$, $B^\dagger_{j,\bk=0}$ superposes a set of plane waves at wavevectors $\brecip_m+\bqorder_j=\brecip_m+\brecip_j/2$. The three sets of plane waves for the three different $j$'s are disjoint and they span the set of plane waves at all the reciprocal lattice vectors of the reduced Brillouin zone, $\bqorder_m$.)

To construct the Bogoliubov Hamiltonian, $H_{\text{Bog}}$, we make the shift in the microscopic Hamiltonian (1) using $b_{m,0}\rightarrow\sqrt{N}\alpha_{m}$.  The Bogoliubov approximation keeps only terms quadratic in $\alpha_m$. These terms are quadratic in the boson operators and can be written in the Nambu form:
\begin{equation}\label{eqn:hambog}
H_{\text{Bog}}=\frac{1}{2}\!\sum_{\bk\in
  \text{BZ}}\!\!\vec{b}_{\bk}^{\dagger}H_{\bk}\vec{b}_{\bk} -\frac{1}{2}\sum_{m\bk} \epsilon_{\bk+\bqorder_m} 
\end{equation}
where
$\vec{b}_{\bk}=(b_{0,\bk},...,b_{m_c,\bk},b_{0,-\bk}^{\dagger},...,b_{m_c,-\bk}^{\dagger})^{T}$. The
Nambu form contains kinetic energy terms of the form $b b^\dagger$ but the original Hamiltonian only has terms of the normal ordered form $b^\dagger b$. The constant term above has been inserted to subtract out an unwanted constant from this rearrangement.

The eigenenergies and eigenvectors of the Bogoliubov quasiparticles are obtained by solving the equation $H_\bk \vec{b}_\bk = E_\bk\sigma_3
\vec{b}_\bk$ where $\sigma_3 =
\text{diag}(1,1,1,\ldots,-1,-1,-1,\ldots)$ is a diagonal
$2m_c\times 2m_c$ matrix.
This is equivalent to a diagonalization using the Bogoliubov
transformation $\vec{b}_{\bk}=T_{\bk} {\bm \beta_{\bk}} $ with
\begin{equation}
T_{\bk}=\left(\begin{array}{cc} u_{\bk} & v_{\bk}\\
    v_{-\bk}^{*} & u_{-\bk}^{*} \end{array}\right)\,,\quad 
u_{\bk} u_{\bk}^\dagger - v_{\bk} v_{\bk}^\dagger=\vec{1}
\end{equation}
giving us a diagonal form of the quadratic Hamiltonian
\begin{equation}\label{eqn:hambog2}
H_{\text{Bog}}= \!\!\!\sum_{\mu\bk} E_{\mu\bk} \beta_{\mu\bk}^\dagger \beta_{\mu\bk}
  + \frac{1}{2}\sum_{\mu\bk}\left(E_{\mu\bk}-\epsilon_{\bqorder_\mu+\bk}\right)\,.
\end{equation}

\begin{figure}
\includegraphics[scale=0.12]{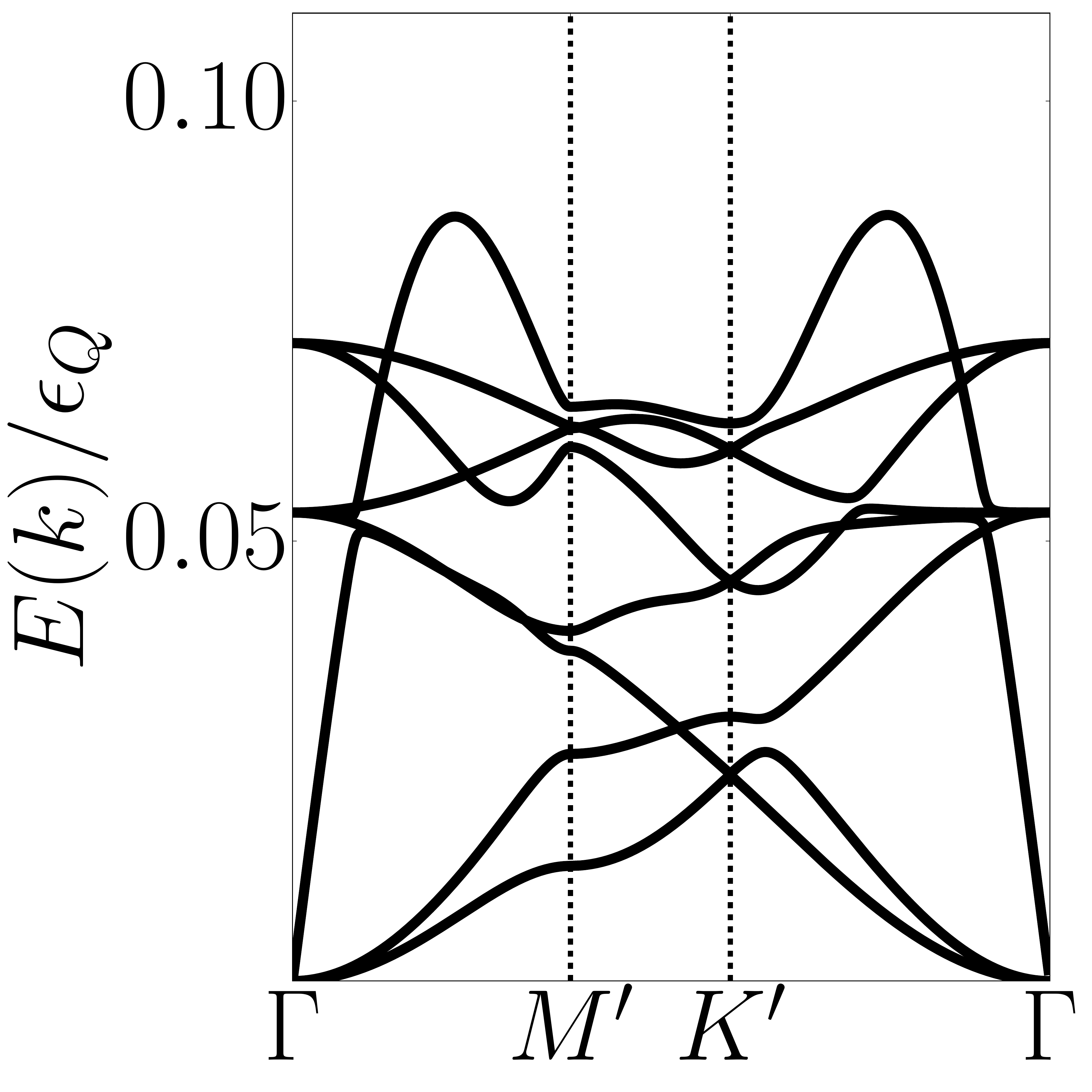}
\includegraphics[scale=0.12]{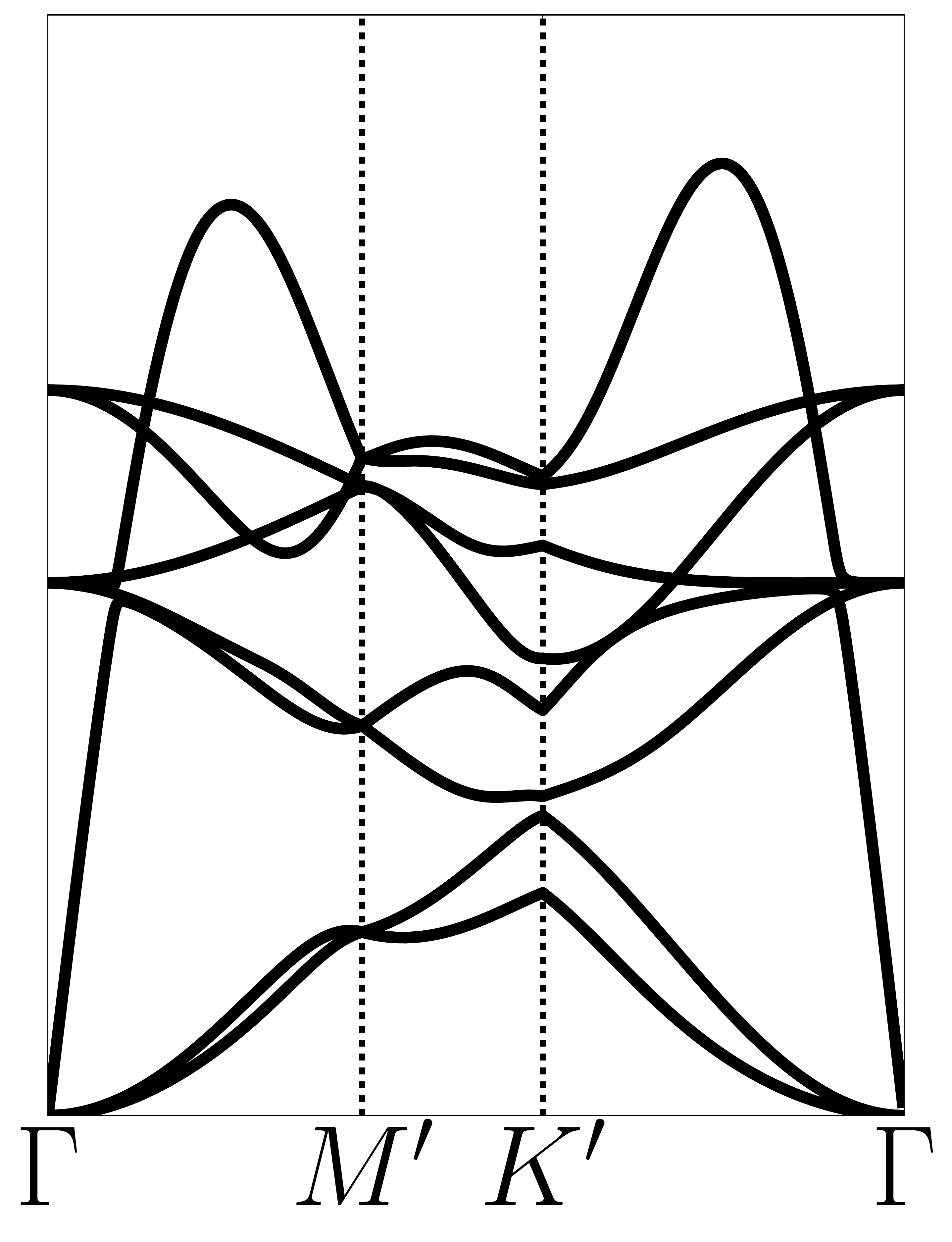}
\caption{\label{spectrum} The eight lowest, positive-energy modes in
  the Bogoliubov spectrum for the state $(c_1,c_2,c_3) =
  (1,1,1)/\sqrt{3}$ (left) $(1,i,0)/\sqrt{2}$ (right) with  $1/QR = 0.3$, $\nbar V_0 =\epq$,  $U=6\epq$. The spectrum converges using a $m_c=271$ plane-wave basis.}
\end{figure}

Figure \ref{spectrum} shows the lowest, positive excitation energies for fluctuations around two coherent states: the symmetric state $(c_1,c_2,c_3) = (1,1,1)/\sqrt{3}$ and the Liu-Wu state $(1,i,0)/\sqrt{2}$. For both states, we find a gapless mode with a linear dispersion.  This corresponds to fluctuations in the overall density, analogous to the usual U(1) superfluid Goldstone mode arising from global phase invariance.  In the next section, we show that the corresponding phase variable is $\delta\Phi = \sqrt{\nbar}\sum_i |c_i|^2\delta\theta_i$ where $\delta\theta_i$ is the fluctuations in the phase of $c_i$.  This variable is not quantized because $0\le |c_i|\le 1$, consistent with the lack of topological defects in the system.

There are also two quadratic modes, $E_{\mu=2,3,\bk}$, which arise
directly as a consequence of the non-Abelian symmetry
generators. Their $k \rightarrow 0$ eigenvectors correspond to
excitations to $p$-band states orthogonal to the condensed
single-particle state. Such excitations are analogous to ``phasons'':
a continuous internal rearrangement of a crystal
\cite{Gopalakrishnan2013} sampling various density configurations
(Fig.~\ref{fig:density}). We note that spin waves in ferromagnets are
also quadratic modes and this dispersion is associated with the order
parameter being a good quantum number of the Hamiltonian. The
analogous conservation law in our system \eqref{eqn:Ham} is the
conservation of the number of particles in the two single-particle
states orthogonal to the condensed state. More mathematically, by
adapting the analysis for the Watanabe-Brauner counting rules
\cite{Watanabe2011,*Watanabe2012a,Takahashi2015,*Nitta2015}, we can
show\footnote{The algorithm is as follows. Suppose we have $n$
broken symmetry generators $P_{1},\ldots,P_{n}$. Define the matrix:
$\Gamma_{ij}= \protect\langle\Psi |[P_{i},P_{j}]|\Psi\protect\rangle$.
The number of modes with even dispersion in $k$ is given by $n_{\text{even}}=\text{rank}(\Gamma)/2$
and the number of odd modes is $n_{\text{odd}}=n-2n_{\text{even}}$.
The broken symmetries of our coherent state are $\Lambda^{4,\ldots,8}$. We
find that $n_{\text{odd}}=1,n_{\text{even}}=2$, agreeing with our
perturbative analysis and numerical simulations.}
 that quadratic modes exist if the expectation values of the
commutators of the non-Abelian SU(3) generators, $\{\Lambda^a\}$, do
not vanish for states within the manifold. This connection with the
enlarged symmetry of the manifold is consistent with these modes
acquiring energy gaps of $(2\Vtilde_1 - \Vtilde_2) \nbar/4$ and
$(\Vtilde_1\Vtilde_2/2)^{1/2}\nbar$ if $\Vtilde_{1,2}\neq 0$ (for the
$(1,i,0)$ state) which we derive in the next section.

We also find Bogoliubov eigenstates with negative energies
corresponding to $s$-band states. Metastability against scattering
into the $s$-band has been addressed \cite{Liu2006,Isacsson2005} and a
metastable $p$-band atomic BEC has been achieved
\cite{Mueller2007,Wirth2011}.


\subsection{Single Band Effective Theory}

Here, we outline a minimal effective theory to describe the long-wavelength excitations of the $p$-band condensate. This will be useful to make a number of analytical statements concerning the Bogoliubov spectrum and the scaling of various thermodynamic quantities. We will use a number-phase representation of the condensate which reveals the physical content of these excitations. (We have set $\hbar=k_B=1$.)

Let $B_{i,\bk}^\dagger$ be the creation operator for a Bloch state with crystal momentum $\bqorder_i+\bk$ near the $M_i$ point with energy $\epik$ ($i=1,2,3$).  We will provide analytic results for the simplified case when the dispersion relation is isotropic $\epik = \epk = \hbar^2\bk^2/2m^*$.  This calculation is easily generalized to the actual anisotropic dispersion but the analytic results are cumbersome.

We study fluctuations around the coherent state (2):
\begin{equation}
\left|\Psi\right\rangle = e^{-N/2} \exp
\left[\sqrt{N}\sum_{i=1}^3 c_i B_{i,\bk=0}^\dagger\right]
\left|\text{vac}\right\rangle.
\end{equation}
In the number-phase representation,
\begin{equation}\label{eqn:numberphase}
  c_j(\br) = \sqrt{n_j(\br)} e^{i\theta_j(\br)}\,,
\end{equation}
this coherent state has a mean-field energy (3) density per unit area of
\begin{multline}
\nbar u_{\text{MF}}=\frac{V_0}{2}(n_1+n_2+n_3)^2 \\
+ \sum_{j=1}^3 \left[\frac{\Vtilde_1}{2}  n_jn_{j+1}[ 1 +\cos(2\theta_j-2\theta_{j+1})]+ \frac{\Vtilde_2}{4}n_j^{2} \right]
\end{multline}
where  $\nbar$ is the mean-field number density, and the addition in the $j$-index is modulo 3. We concentrate on long-wavelength fluctuations ($|\bk|\ll \qorder$) in the amplitudes
\begin{equation}
  \sqrt{N} c_j \to \sqrt{N} c_j(\br) =  \sqrt{\frac{N}{L^2}}
  \sum_\bk^{k\ll \qorder} c_{j,\bk} e^{i\bk\cdot\br}
\end{equation}
where $L^2$ is the area of the system.
To be more precise, we consider states of the form
\begin{equation}
  \begin{split}
    \left|\{c_j(\br)\}\right\rangle &=
    e^{-N/2}\exp\left[\sqrt{N}\int\!\! c_j(\br)\psi^\dagger_j(\br)\, d^2\br\right]\left|\text{vac}\right\rangle \\
    &=\exp\left[\sqrt{N}\sum_{j,\bk}c_{j\bk}B_{j,\bk}^\dagger\right]\left|\text{vac}\right\rangle\,,\\
    c_j(\br) &= \frac{1}{L} \sum_{j,\bk} c_{j,\bk}  e^{i\bk\cdot\br}\,,\quad
      \psi^\dagger_j(\br) = \frac{1}{L}\sum_\bk B^\dagger_{j,\bk} e^{-i\bk\cdot\br}
  \end{split}
\end{equation}
where $c_{j,\bk}$ is not small only for $k\ll Q$ and $\psi_j(\br)$ is the field operator projected onto the Bloch states in the $p$-band around the $M_j$ point.  The Lagrangian density ${\cal L}$ for the long-wavelength fluctuations can be written as
\begin{equation}
  {\cal L} = \nbar \sum_{j=1}^3 c^*_j \left(i\hbar\partial_t -\epsilon_{j\hat\bk}\right) c_j - \nbar u_{\text{MF}}[\{c_j(\br)\}]
\end{equation}
where $\epsilon_{j\hat\bk}$ is obtained from the single-particle band energies $\epjk$ by replacing $\bk\to \hat\bk = -i\hbar\nabla$.

For small fluctuations in the density and phase, $n_j(\br) = \nbar_j+\dn_j$ and $\theta_j = \thetabar_j + \dtheta_j$, we write $c_j \simeq \sqrt{\nbar_j} \exp(i\thetabar_j) ( 1 + \dn_j/2\nbar_j+ i \dtheta_j)$ where $\sqrt{\nbar_j}\exp(i\thetabar_j)$ is the mean field value of $c_j$ that minimizes $u_{\text{MF}}$.  Then, we expand ${\cal L}$ and collect the terms quadratic in $\delta\theta_j$ and $\delta n_j$. Consider first the SU(3) symmetric Hamiltonian with $\Vtilde_{1,2}=0$ with the $S^5$ manifold of degenerate coherent states described by any $(c_1,c_2,c_3)$ with $\nbar = \nbar_1+\nbar_2+\nbar_3$ fixed. The quadratic fluctuations are described by the Lagrangian density:
\begin{multline}
  \delta {\cal L}_{S^5} = \sum_{i=1}^3 \bigg[-\dn_i\partial_t\dtheta_i  - \frac{V_0}{2}(\dn_1+\dn_2+\dn_3)^2\\
    -\frac{1}{2m^*}\left(\nbar_i|\nabla \dtheta_i|^2+ \frac{1}{4\nbar_i}|\nabla \dn_i|^2\right)\bigg] \,.
\end{multline}
Note that $\dn_i$ and $\dtheta_i$ are canonically conjugate variables. Three pairs of natural canonical conjugates for this problem are
\begin{multline}\label{eq:newbasis}
\nu_i  = \sum_j P_{ij} \frac{\dn_j}{\sqrt{\nbar_j}}\,,\quad
\phi_i  = \sum_j P_{ij} \sqrt{\nbar_j}\dtheta_j\,,\quad \\
 P = \begin{pmatrix}
    \sqrt{\frac{\nbar_1}{\nbar}} &\sqrt{\frac{\nbar_2}{\nbar}}&\sqrt{\frac{\nbar_3}{\nbar}}\\
   -\sqrt{\frac{\nbar_2}{\nbar-\nbar_3}}&\sqrt{\frac{\nbar_1}{\nbar-\nbar_3}}&0\\
    -\sqrt{\frac{\nbar_1\nbar_3}{\nbar(\nbar-\nbar_3)}} &-\sqrt{\frac{\nbar_2\nbar_3}{\nbar(\nbar-\nbar_3)}}&\sqrt{\frac{\nbar-\nbar_3}{\nbar}}
  \end{pmatrix}
\end{multline}
Using these canonical variables, we can write
\begin{equation}
\delta L_{S^5} = \sum_{i=1}^3 \left[-\nu_i\partial_t\phi_i -
    \frac{|\nabla \phi_i|^2}{2m^*}+ \frac{|\nabla \nu_i|^2}{8m^*}\right]
  - \frac{V_0\nbar}{2} \nu_1^2
\end{equation}
The spectrum for this system can be easily extracted by comparing this with the Langrangian for a simple harmonic oscillator with frequency $\omega$: $L_{\text{SHO}} = p\dot q- p^2/2m - m\omega^2q^2/2$. We find three gapless modes. Mode 1 has the dispersion relation $E_{1\bk} =\sqrt{\epk(\epk + 2V_0\nbar)}$ which is linear in the wavevector $k$ for small $k$. This corresponds to overall density fluctuations $\delta n = \sqrt{\nbar}\nu_1 = \dn_1+\dn_2+\dn_3$. The conjugate phase variable is $\delta\Phi = \phi_1/\sqrt{\nbar} = \nbar_1\dtheta_1+\nbar_2\dtheta_2+\nbar_3\dtheta_3$. The two other modes are degenerate and simply have the non-interacting dispersion $E_{2\bk}=E_{3\bk}=\epk$. In second-quantized form, the annihilation operators for the three modes are
\begin{align}\label{eq:bogcreate}
  a_{j\bk} &= l_{j\bk} \phi_{j\bk} - \frac{i}{2l_{j\bk}}\nu_{j\bk}\,,
  \\ \quad
  l_{1\bk}^2 &=
  \left(\frac{\epk}{2V_0\nbar+\epk}\right)^{1/2}\!\! =
  \frac{\epk}{E_1(\bk)}\,,\qquad
  l_{2/3,\bk}= 1\,. \notag
\end{align}
For the $j=1$ mode which is linear at small $k$, we see that
$l_{1\bk}^2$ scales as $k$ at small $k$.

We can show that the existence of a
linear mode and two quadratic modes is robust when we restore the
anisotropy in the band energies around the three $M$ points. The
quadratic modes become non-degenerate and their energies do depend on
the interaction strength $V_0$. Moreover, the spectrum becomes
dependent on the choice of the mean-field coherent state.

The $\Vtilde_{1,2}$ interaction terms break the $S^5$ symmetry. The ground state is $(c_1,c_2,c_3)=(1,\pm i,0)/\sqrt{2}$ with a U(1) symmetry for the overall phase. Fluctuations around this state can be described by the number and phase fluctuations at the two condensed amplitudes $c_{1,2}$ and a decoupled single-particle Hamiltonian for fluctuations around $c_3=0$.
\begin{align}
    \delta {\cal L}_2 &= -\left[\nu_1\partial_t\phi_1 
                        -\phi_1 \ephatk \phi_1+\frac{\nu_1}{8}\left(4V_0\nbar+\Vtilde_2\nbar + 2\ephatk\right)\nu_1\right]\notag\\
  -&\left[\nu_2\partial_t\phi_2 +\phi_2\left(\Vtilde_1\nbar+\ephatk\right)\phi_2
    + \frac{\nu_2}{8}\left(\Vtilde_2\nbar+2\ephatk\right)\nu_2\right]\,,\notag\\
  \delta {\cal L}_3 &=  \nbar c^*_3 \left(i\partial_t - \ephatk-
               \frac{\Vtilde_1\nbar}{2}+\mu-V_0\nbar
               \right) c_3
\end{align}
with the chemical potential $\mu = 2\nbar u_{\text{MF}}= (V_0-\Vtilde_2/4)\nbar$. Mode 1 for overall density fluctuations remains linear. The quadratic modes from the SU(3)-symmetric case now have energy gaps of $(\Vtilde_1\Vtilde_2/2)^{1/2}\nbar$ and $(2\Vtilde_1-\Vtilde_2)\nbar/4$.

\section{Finite Size Scaling}
\label{sec:scaling}

We found in section \ref{sec:singleparticle} that a coherent state of the form \eqref{eqn:manifold} minimizes the energy in the $p$-band of the single-particle band structure. In this section, we use the excitation spectrum derived in the previous section to perform a stability analysis of this coherent state. By computing the condensate depletion and the normal fluid densiy, we verify our assertion in section \ref{sec:nobkt} that, while the system is condensed at zero temperature, both the condensate depletion and normal fluid density grows with system size at any non-zero temperature. Thus, the system is neither condensed nor superfluid in the thermodynamic limit at any non-zero temperature.

\textit{Condensate depletion.} The condensate depletion $\Delta$ is defined as the fraction of particles with momenta different from the ones in the coherent state (2). 
\begin{equation}\label{depletion}
  \Delta = \frac{1}{N}\sum_{j,\bk\neq 0}\langle c_{j,\bk}^\dagger
  c_{j,\bk}  \rangle
\end{equation}
At the level of our approximation of small fluctuations,
\begin{align}\label{depletionnumberphase}
  \Delta &\simeq \frac{1}{N}\sum_{j,\bk\neq 0}\left\langle
    \left(\frac{\nu_{j,\bk}}{2}- i \phi_{j,\bk}\right)
    \left(\frac{\nu_{j,-\bk}}{2}+ i \phi_{j,-\bk}\right)
  \right\rangle\notag \\
  &= \frac{1}{2\nbar}\sum_j\int\!\!\frac{d^2\bk}{(2\pi)^2}\!\!\left[\frac{l_{j\bk}^2+l_{j\bk}^{-2}-2}{2}
  + (l_{j\bk}^2+l_{j\bk}^{-2})N_{j,\bk}\right]
\end{align}
where $N_{i\bk} = 1/(e^{E_{j\bk}/T}-1)$ is the Bose occupation number
of the eigenstate with energy $E_{i,\bk}$.
The first term is the depletion at zero temperature. From the
small-$k$ behavior of quasiparticle dispersion relations,
we can see that the integrand of the first term is dominated by the
linear mode, as found in a conventional U(1) superfluid. We can check
that it is finite in 2D.
This is consistent with the fact that Bose condensation is possible \emph{at} zero
temperature in two dimensions.
The second term arises from the thermal excitation of
quasiparticles. The temperature scaling of this term depends also solely on
the form of the power law in the quasiparticle dispersion relations.
At any given low temperature $T$, the contribution from 
long-wavelength fluctuations dominate the integrand. The temperature
dependence can be obtained by noting that 
$N_{i\bk}\simeq T/E_{i\bk}$ for $E_{i\bk}\ll T$ and summing only up to $E\sim T$. 
It can be shown that both linear and quadratic modes contribute terms
of the form $T/k^2$ in the integrand. Thus, in two dimensions, the thermal depletion
scales as $T\log(LT)$ where $L$ is the linear size of the system. We
note that this temperature scaling depends solely on the
form of the power law in the dispersion relation of the different
quasiparticle modes. 

\textit{Superfluid density.} The local current density is given by $\bJ = N \sum_i c_i^* \nabla_{\hat\bk}\ephatk c_i$. For an isotropic quadratic dispersion around the $M$ points, this gives $\bJ = -i N \sum_i (c_i^*\nabla_\br c_i-c_i\nabla_\br c_i^*)/2m^*$. If we confine our attention to slow spatial variations only, the current is given in the number-phase representation by 
\begin{equation}
  \bJ \simeq  \frac{1}{m^*}\sum_i n_i \nabla \theta_i \simeq \frac{1}{m^*}\sum_i
  \left(\nbar_i \nabla \dtheta_i + \dn_i \nabla \dtheta_i\right)\,.
\end{equation}
The first term involves excitations of a single quasiparticle while the latter involves two quasiparticles. The first is longitudinal and therefore does not contribute to the normal fluid response. The second term is diagonal in the index $i$ and remains so after the orthogonal basis transformation \eqref{eq:newbasis}. Its Fourier transform is
\begin{align}
  \bJ_{\perp\bq} &\simeq  
  \frac{1}{m^*}\sum_i \int \nu_i (\nabla\phi_i) _{\perp\bq}  e^{-i\bq\cdot\br}\frac{d^2\br}{L^2} \notag\\
  &= \frac{i}{m^*}\sum_{i\bk}\bk _{\perp\bq}\,\nu_{i,\bq-\bk} \phi_{i\bk} \\
  &=\frac{1}{2m^*}\sum_{i\bk}\frac{l_{i,\bq-\bk}}{l_{i\bk}}\bk _{\perp\bq}\,  (a_{i\bq-\bk}-a_{i\bk-\bq}^\dagger) (a_{i\bk}+a_{i-\bk}^\dagger)\,. \notag
\end{align}
The normal fluid density is given by \cite{nozierespines2}
\begin{equation}\label{nozieres}
\rho_n = \frac{2}{L^2 Z} \lim_{\bq \to 0}\sum_{\nu\nu'} 
\frac{e^{-\mathcal{E}_\nu/T}|\langle \nu'| J_{\perp\bq} |\nu\rangle|^2}{\mathcal{E}_{\nu'}-\mathcal{E}_\nu}
\end{equation}
where $\nu$ labels eigenstates with energies $\mathcal{E}_\nu$ of the gas of Bogoliubov excitations with partition function $Z$ and $J_{\perp\bq}$ is the component of the current operator transverse to $\bq$. Inserting the quasiparticle spectrum gives
\begin{equation}
  \begin{split}
    \rho_n 
&= \frac{2}{m^{*2}L^2} \lim_{\bq \to 0}\sum_{i\bk} \left(\frac{\bk _{\perp\bq} l_{i,\bq-\bk}}{l_{i\bk}}\right)^2 \\
&\qquad\times\left[\frac{N_{i,\bq-\bk} +N_{i\bk}}{E_{i,\bq-\bk} +E_{i\bk}}-\frac{N_{i,\bq-\bk} -N_{i\bk}}{E_{i,\bq-\bk} -E_{i\bk}}
\right]\\
&\simeq \frac{4T}{m^{*2}L^2} \lim_{\bq \to 0}\sum^{E_{i\bk}<T}_{i\bk}
\left(\frac{\bk _{\perp\bq}}{E_{i\bk}}\right)^2\,.
\end{split}
\end{equation}
using $N_{i\bk}\simeq T/E_{i\bk}$ and summing up to $E\sim T$.
Again, the temperature scaling of this quantity depends solely on the
form of the power law in the dispersion relation of the different
quasiparticle modes.  The contribution from the linear mode gives a
dependence of $T^3$ while the quadratic mode gives $T\log(LT)$. The
first term is found in the conventional 2D superfluid. The second
term, which dominates at low temperatures, is a novel feature arising
from the extra conservation laws that gave rise to the quadratic
modes, as discussed at the end of section
\ref{subsec:numerical}. We
should point that that this result does not depend on the precise form
of the interaction as long as we can make the approximation that
$\Vtilde_{1,2}$ is small compared to the thermal energy.

When the anisotropy of the dispersion around the $M$ points is included in the calculation, these temperature dependences are robust. There is also a reduction of the superfluid fraction from unity at zero temperature, as expected on general grounds due to the loss of Galilean invariance.

In summary, we have computed the finite-size scaling
behaviour of our $S^5$ condensate. Both the condensate depletion and
the normal fluid density scales as $T\log(TL)$. At any non-zero
temperature, these quantities diverge as the system size $L$ grows. On
the other hand these quantities vanishes at $T=0$. This means that the
condensate is stable at zero temperature against quantum
fluctuations. However, thermal fluctuations destroy condensation
(Mermin-Wagner-Hohenberg theorem) and superfluidity in the
thermodynamic limit due to the thermal excitation of gapless
quasiparticles.

\section{order-by-disorder splitting}
\label{sec:orderbydisorder}

The $S^{5}$ symmetry is emergent meaning that the generators described above do not commute with the Hamiltonian but only do so in the expectation value of the macroscopic coherent state. Such emergent symmetries are typically broken by the ``order-by-disorder'' mechanism \cite{Song2007,Turner2007,Barnett2012} which reduces the symmetry by picking the state that minimizes the quantum zero-point energy $E_{\text{zp}}=\frac{1}{2}\sum_{\mu\bk} (E_{\mu\bk}- E^{(0)}_{\mu\bk})$ where $E^{(0)}_{\mu\bk}$ are the non-interacting band energies. In systems where order-by-disorder is typically important, this quantity is on the order of $\epq$ \cite{Barnett2012}. We have evaluated the zero-point energy for the parameters considered in Fig.~\ref{spectrum}. We find that the order-by-disorder mechanism favors the symmetrically condensed state, $(c_1,c_2,c_3)\propto (1,1,1)$ and its three other degenerate counterparts: $(-1,1,1)$ and permutations. In other words, this reduces the degeneracy on the $S^5$ manifold to a U(1)$\otimes\mathbb{Z}_4$ symmetry. These states have a zero-point energy per particle of $ E_{\text{zp}}/N \simeq 4 \times 10^{-2}\epq /\ncell$ where $\ncell=\sqrt{3}\nbar a^2/2$ is the number of particles per unit cell.  (The $p$-band has a bandwidth of $\sim 10^{-1}\epq$ for these parameters.) However, the range of zero-point energies over the whole manifold is only $1\%$ of this quantity: $\Delta E_{\text{zp}}/N \sim 10^{-4} \epq/\ncell$, with the $(1,i,0)$ state having the highest zero-point energy. We believe that this surprisingly small range can be related to the small matrix elements for the intervalley superexchange contribution to the zero-point energy (Fig.~\ref{fig:exchange} right).  This involves intermediate states produced by momentum transfers of the order of $Q/2$ from a condensed wavevector $M$, and we see numerically that the splitting scales approximately as $V_{Q/2}^2/V_0$ for $V_0 \gg V_{Q/2}$ ($= 6\times 10^{-2}V_0$ in Fig.~\ref{spectrum}).  Such small energy differences between the coherent states in the $S^5$ manifold means that they should all be accessible at low temperatures.

\section{Suppression of the BKT transition}
\label{sec:bktsuppress}

We return now to the issue of the small energy gap due to a small
$V_Q$: $\Egap \simeq 10^{-5}\epq$ for parameters in
Fig.~\ref{spectrum}. This reflects the anisotropy on the $S^5$
manifold of coherent states, reducing the symmetry to U(1). We
estimate the temperature scale at which the breaking of the symmetry
of the $S^5$ degenerate manifold due to a small non-zero interaction
$V_Q$ which couples bosons at two $M$ points by a momentum transfer of
$Q$. This reduces the symmetry of the degenerate manifold to U(1).

We borrow from Nelson and Pelcovits \cite{NelsonPelcovits1977} and Fellows \emph{et al} \cite{Fellows2012} and consider the O($M$+2) non-linear sigma model with a small anisotropic term, defined by the $(M+2)\times(M+2)$ matrix $D$, that breaks the symmetry to an O(2) model. This is described by the energy density:
\begin{equation}
  {\cal H} =  \frac{J}{2}
  (\nabla\vec{n})^2+\frac{\Jperp }{2a^2} \vec{n}^{\text{T}}D \vec{n}
\end{equation}
where $\vec{n}$ is a unit vector on the $S^{M+1}$-sphere, $\Jperp$ is a dimensionless measure of the anisotropy and $a = 2\pi/\sqrt{3} Q$ is the lattice spacing. In the absence of the anistropy, the Mermin-Wagner theorem states that the system is disordered in the thermodynamic limit in two dimensions at any non-zero temperature. For a non-zero $\Jperp \ll 1$, a BKT transition occurs at a critical temperature $T_c\simeq J/\ln(J/\Jperp)$. In the opposite limit of large $\Jperp$, this is equivalent to the O(2) model which has a critical temperature of  $T_{\text{BKT}} \simeq \pi J/2$.

We should also note that the anisotropy gives rise to topologically stable vortices. The size of these vortices diverge as $\xi \sim a/\sqrt{J/\Jperp}$ as $\Jperp \to 0$.

Our system cannot be mapped directly onto an O($M$+2) model. However, we believe that we can use these results to estimate the effect of anistropy. We estimate that $J \sim \hbar^2\nbar/2m^*$ where $m^*$ is the effective mass of the single-particle dispersion relation around the $M$ points. Since the anisotropy arises from intervalley exchange, the anistropy energy per unit area is controlled by $\nbar^2 \Vtilde_1\sim \nbar \Egap$. We estimate $\Jperp/ a^2 \sim \nbar \Egap $. This gives $\Jperp/J \sim
8\pi^2m^*\Egap/\hbar^2= (4\pi^2/3)(m^*/m)(\Egap/\epq)$.

For Fig.~3 where $U = 6\epq$, $V_0 = \nbar\epq$ and $\Vtilde_1\simeq 10^{-5} V_0$, we find  $m^*/m \sim 0.2$ this gives $\Jperp/J \sim  10^{-5}$. So, the BKT transition temperature is suppressed by a factor of $1/\ln(\Jperp/J) \sim 10^{-1}$ for an infinite system. This will be observable for systems larger than the vortex size $\xi \sim 300 a$.  This leaves us scope to explore the non-Abelian condensate as a failed superfluid.

\section{Conclusions and outlook}
\label{sec:conclude}

In this work, we have proposed a non-Abelian condensate with spatial
density modulations in two dimensions. We have demonstrated that this is a local minimum of the energy at zero temperature. We argued that the loss of topological vortices means that we do not expect superfluidity in the system at non-zero temperatures. This is verified by a calculation of how condensation and superfluidity is lost as the system is increased at non-zero temperatures. 

Can this non-Abelian condensate be a candidate for a ``supersolid'' phase that spontaneously breaks both translational and global gauge symmetries? The condensation at non-zero momenta may be induced by certain two-body interaction potentials with negative Fourier components at the ordering wavevector.  Such condensation may occur even in the absence of an applied field by creating a roton instability \cite{Henkel2010,Macri2013}. Such a system is generically an Abelian condensate with decoupled Bogoliubov and phonon modes. (In the context of our $S^5$ manifold, the interactions creating the roton instability will generically determine all the relative weights and phases of the amplitudes $c_i$, other than the ones responsible for these U(1) modes.) Therefore, one expects to see a BKT transition in contrast to \cite{Nyeki2017}. Nevertheless, we can show \footnote{In preparation.} that non-Abelian condensates can be local minima in mean-field theory for special fine-tuned Hamiltonians. These states, however, appear to be dynamically unstable in general, as evidenced by imaginary eigenvalues in their Bogoliubov spectra. A notable exception arises if the single-particle spectrum deviates from the typical quadratic kinetic energy dispersion (\emph{e.g.}~due to band structure, or internal degrees of freedom) where we have found a condensate with SU(2) symmetry in addition to the U(1) translational symmetries. Such a setup will be the focus of future studies. 

In summary, we have proposed a scenario for a condensate with non-Abelian features, such as a lack of a BKT transition and additional gapless ``phason" modes. We believe this is the first example of such a condensate that exploits spatial structure instead of additional internal degrees of freedom. By leveraging the single-particle degeneracy of the $p$-band, we study an interaction that does not spoil the SU(3) symmetry of the system at the mean-field level. We find a ``failed superfluid'' in two dimensions. Our scenario is not confined to a triangular lattice and is anticipated to generalize to degenerate higher band condensates in \emph{e.g.}~square, hexagonal lattices and in three dimensions. Intriguingly, our failed superfluid shares similar low-temperature behavior and a lack of a BKT transition with the $^4$He bilayer on graphite \cite{Nyeki2017}. Nevertheless, a complete theory for this helium system that motivated our story remains elusive.

SL is supported by the Imperial College President's Scholarship.  PC is supported by the US Department of Energy, Office of Basic Energy Sciences grant DE-FG02-99ER45790. Since the submission of this work, an SO(3) generalization of the spontaneous optical lattice experiment has been proposed \cite{RodriguezChiacchio2018}.

\begin{appendix}
\section{Dipolar interactions}
\label{app:dipolar}

In this section, we review the work of Fischer~\cite{Fischer2006}
which demonstrated how a finite-range interaction can be realized for dipolar bosons confined to a cloud in the $xy$-plane.

Consider dipolar bosons of mass $m$ with dipole moment $d_e$, polarized by a
strong electric field in the $z$-direction. The interaction between
two bosons at a (three-dimensional) displacement of $\vec{r}$ consists
of two components. Firstly, there is a contact interaction
parametrized by an $s$-wave scattering length $a_s$ or an interaction
strength $g_{\text{3D}} = 4\pi \hbar^2a_s/m$. There is also a
dipole-dipole interaction of the form $V_{dd}(\vec{r}) = (3g_d/4\pi
r^3)(1-3z^2/r^2)$ with $g_d = d_e^2/3\epsilon_0$.  This is repulsive
when $x^2+y^2 \gg z^2$ and attractive when $x^2+y^2 \ll z^2$ (when the dipoles are
nearly collinear in the $z$-direction).

When these bosons are confined by harmonic trap to a Gaussian wavepacket of width $d_z$
in the $z$-direction, the Fourier transform of the effective
interaction in the 2D plane can be written as
\begin{equation}
\begin{split}
V_{\bq}&= \frac{g_d}{d_z}\left[
  \frac{1+g_{\text{3D}}/2g_d}{\sqrt{\pi/2}} - \frac{3}{2} q d_z
  w\left(\frac{qd_{z}}{\sqrt{2}}\right)    \right]\\
&\qquad\text{with\ }
   \quad w(x)= e^{x^2}\text{erfc}(x)\,,
\end{split}
\end{equation}
where $\bq$ is the 2D wavevector of the Fourier transform.
Fischer \cite{Fischer2006} proposed that the contact interaction
strength $g_{\text{3D}}$ can be tuned
to be equal to $g_d$ so that $V_\bq \to 0$ as $q\to \infty$. Thus, the
short-range contributions from the dipolar interaction and contact
interaction cancel each other, producing an interaction with
lengthscale $d_z$. 
For this study, we want this lengthscale to be large compared to the
wavelength $\sim 1/Q$ of the density modulations of our coherent state
\eqref{eqn:manifold}. This corresponds to the condition that
confinement in the $z$-direction must be larger than $\sqrt{3}a$ where
$a$ is the length of the triangular lattice vector.  This suppresses
intervalley processes that break the $S^5$ symmetry.

\begin{figure}[bht]
\includegraphics[scale=0.12]{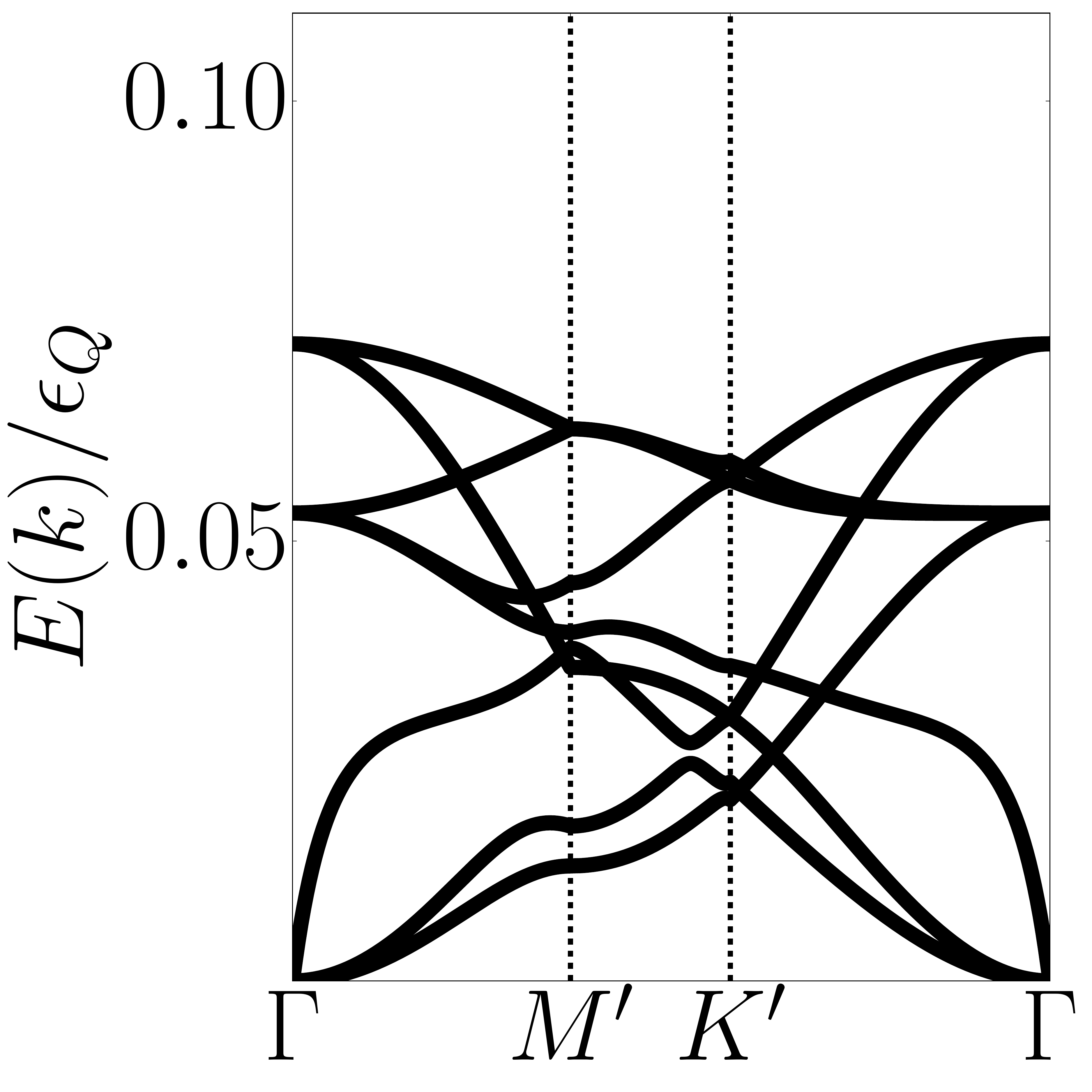}
\includegraphics[scale=0.12]{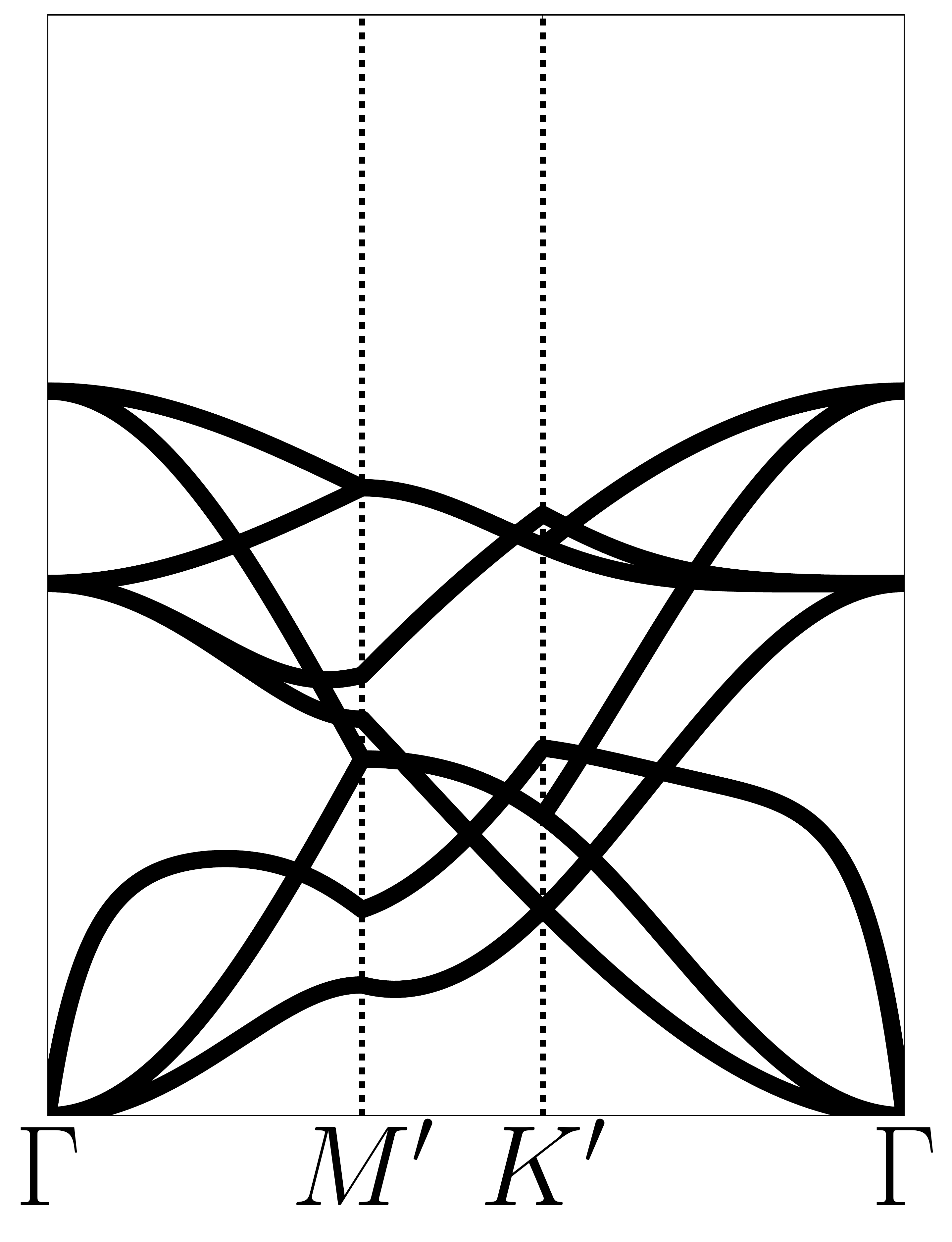}
\caption{\label{dipspectrum} The eight lowest, positive-energy modes
  in the Bogoliubov spectrum for the state $c_1=c_2=c_3$ (left) and
  $c_1=1, c_2=c_3=0$ (right). Parameters:
  $g_{d}=g_{\text{3D}}$, $d_z = 7.5/Q$,
  $3\sqrt{2/\pi}\,g_d\nbar/2d_z = \epq$, $U=6\epq$.}
\end{figure}

In summary, we impose two conditions on the interaction to observe the $S^5$ symmetry
\begin{equation}\label{dipolecondition}
g_{d}=g_{\text{3D}} \,,\qquad d_{z}>\sqrt{3} a\,,
\end{equation}
which can be achieved by using a Feshbach resonance and by adjusting
the out-of-plane confinement of the trap. We plot the resulting
Bogoliubov spectrum in Fig.~\ref{dipspectrum}.

\end{appendix}
 We should recall that the bosons are regarded as
quasi-2D because they are condensed
in the ground state of the confinement potential in the
$z$-direction. The next lowest state is higher in energy by
$\hbar^2/md^2_z$. We should not allow $d_z$ to be so large that $\hbar^2/md_z^2$
becomes smaller than the bandwidth of the $p$-band for motion in the
$xy$-plane. This can be achieved without violating the condition
\eqref{dipolecondition} in a deep optical lattice where the bandwidth is only a
fraction of $\epq$.  For instance, $U=6\epq$ (used
in Fig.~\ref{spectrum}), gives a bandwidth of the order of
$0.05\epq$. Thus, there is a window of $d_z$ where the
system remains quasi-two-dimensional while the interparticle interaction is smooth
over the lattice spacing.

\bibliography{pOrb} 

\begin{thebibliography}{35}%
\makeatletter
\providecommand \@ifxundefined [1]{%
 \@ifx{#1\undefined}
}%
\providecommand \@ifnum [1]{%
 \ifnum #1\expandafter \@firstoftwo
 \else \expandafter \@secondoftwo
 \fi
}%
\providecommand \@ifx [1]{%
 \ifx #1\expandafter \@firstoftwo
 \else \expandafter \@secondoftwo
 \fi
}%
\providecommand \natexlab [1]{#1}%
\providecommand \enquote  [1]{``#1''}%
\providecommand \bibnamefont  [1]{#1}%
\providecommand \bibfnamefont [1]{#1}%
\providecommand \citenamefont [1]{#1}%
\providecommand \href@noop [0]{\@secondoftwo}%
\providecommand \href [0]{\begingroup \@sanitize@url \@href}%
\providecommand \@href[1]{\@@startlink{#1}\@@href}%
\providecommand \@@href[1]{\endgroup#1\@@endlink}%
\providecommand \@sanitize@url [0]{\catcode `\\12\catcode `\$12\catcode
  `\&12\catcode `\#12\catcode `\^12\catcode `\_12\catcode `\%12\relax}%
\providecommand \@@startlink[1]{}%
\providecommand \@@endlink[0]{}%
\providecommand \url  [0]{\begingroup\@sanitize@url \@url }%
\providecommand \@url [1]{\endgroup\@href {#1}{\urlprefix }}%
\providecommand \urlprefix  [0]{URL }%
\providecommand \Eprint [0]{\href }%
\providecommand \doibase [0]{http://dx.doi.org/}%
\providecommand \selectlanguage [0]{\@gobble}%
\providecommand \bibinfo  [0]{\@secondoftwo}%
\providecommand \bibfield  [0]{\@secondoftwo}%
\providecommand \translation [1]{[#1]}%
\providecommand \BibitemOpen [0]{}%
\providecommand \bibitemStop [0]{}%
\providecommand \bibitemNoStop [0]{.\EOS\space}%
\providecommand \EOS [0]{\spacefactor3000\relax}%
\providecommand \BibitemShut  [1]{\csname bibitem#1\endcsname}%
\let\auto@bib@innerbib\@empty
\bibitem [{\citenamefont {Ny{\'{e}}ki}\ \emph {et~al.}(2017)\citenamefont
  {Ny{\'{e}}ki}, \citenamefont {Phillis}, \citenamefont {Ho}, \citenamefont
  {Lee}, \citenamefont {Coleman}, \citenamefont {Parpia}, \citenamefont
  {Cowan},\ and\ \citenamefont {Saunders}}]{Nyeki2017}%
  \BibitemOpen
  \bibfield  {author} {\bibinfo {author} {\bibfnamefont {J.}~\bibnamefont
  {Ny{\'{e}}ki}}, \bibinfo {author} {\bibfnamefont {A.}~\bibnamefont
  {Phillis}}, \bibinfo {author} {\bibfnamefont {A.}~\bibnamefont {Ho}},
  \bibinfo {author} {\bibfnamefont {D.}~\bibnamefont {Lee}}, \bibinfo {author}
  {\bibfnamefont {P.}~\bibnamefont {Coleman}}, \bibinfo {author} {\bibfnamefont
  {J.}~\bibnamefont {Parpia}}, \bibinfo {author} {\bibfnamefont
  {B.}~\bibnamefont {Cowan}}, \ and\ \bibinfo {author} {\bibfnamefont
  {J.}~\bibnamefont {Saunders}},\ }\href {\doibase 10.1038/nphys4023}
  {\bibfield  {journal} {\bibinfo  {journal} {Nat. Phys.}\ }\textbf {\bibinfo
  {volume} {13}},\ \bibinfo {pages} {455} (\bibinfo {year} {2017})}\BibitemShut
  {NoStop}%
\bibitem [{\citenamefont {Berezinskii}(1971)}]{Berezinskii1971}%
  \BibitemOpen
  \bibfield  {author} {\bibinfo {author} {\bibfnamefont {V.~L.}\ \bibnamefont
  {Berezinskii}},\ }\href {\doibase 10.1051/jp3:1993206} {\bibfield  {journal}
  {\bibinfo  {journal} {Sov. Phys. JETP}\ }\textbf {\bibinfo {volume} {32}},\
  \bibinfo {pages} {493} (\bibinfo {year} {1971})}\BibitemShut {NoStop}%
\bibitem [{\citenamefont {Kosterlitz}\ and\ \citenamefont
  {Thouless}(1973)}]{Kosterlitz1973}%
  \BibitemOpen
  \bibfield  {author} {\bibinfo {author} {\bibfnamefont {J.~M.}\ \bibnamefont
  {Kosterlitz}}\ and\ \bibinfo {author} {\bibfnamefont {D.~J.}\ \bibnamefont
  {Thouless}},\ }\href {http://stacks.iop.org/0022-3719/6/i=7/a=010} {\bibfield
   {journal} {\bibinfo  {journal} {J.~Phys.~C: Solid State Physics}\ }\textbf
  {\bibinfo {volume} {6}},\ \bibinfo {pages} {1181} (\bibinfo {year}
  {1973})}\BibitemShut {NoStop}%
\bibitem [{\citenamefont {Landig}\ \emph {et~al.}(2016)\citenamefont {Landig},
  \citenamefont {Hruby}, \citenamefont {Dogra}, \citenamefont {Landini},
  \citenamefont {Mottl}, \citenamefont {Donner},\ and\ \citenamefont
  {Esslinger}}]{Landig2016}%
  \BibitemOpen
  \bibfield  {author} {\bibinfo {author} {\bibfnamefont {R.}~\bibnamefont
  {Landig}}, \bibinfo {author} {\bibfnamefont {L.}~\bibnamefont {Hruby}},
  \bibinfo {author} {\bibfnamefont {N.}~\bibnamefont {Dogra}}, \bibinfo
  {author} {\bibfnamefont {M.}~\bibnamefont {Landini}}, \bibinfo {author}
  {\bibfnamefont {R.}~\bibnamefont {Mottl}}, \bibinfo {author} {\bibfnamefont
  {T.}~\bibnamefont {Donner}}, \ and\ \bibinfo {author} {\bibfnamefont
  {T.}~\bibnamefont {Esslinger}},\ }\href {\doibase 10.1038/nature17409}
  {\bibfield  {journal} {\bibinfo  {journal} {Nature}\ }\textbf {\bibinfo
  {volume} {532}},\ \bibinfo {pages} {476} (\bibinfo {year}
  {2016})}\BibitemShut {NoStop}%
\bibitem [{\citenamefont {L{\'{e}}onard}\ \emph {et~al.}(2016)\citenamefont
  {L{\'{e}}onard}, \citenamefont {Morales}, \citenamefont {Zupancic},
  \citenamefont {Esslinger},\ and\ \citenamefont {Donner}}]{Leonard2016}%
  \BibitemOpen
  \bibfield  {author} {\bibinfo {author} {\bibfnamefont {J.}~\bibnamefont
  {L{\'{e}}onard}}, \bibinfo {author} {\bibfnamefont {A.}~\bibnamefont
  {Morales}}, \bibinfo {author} {\bibfnamefont {P.}~\bibnamefont {Zupancic}},
  \bibinfo {author} {\bibfnamefont {T.}~\bibnamefont {Esslinger}}, \ and\
  \bibinfo {author} {\bibfnamefont {T.}~\bibnamefont {Donner}},\ }\href
  {http://www.nature.com/doifinder/10.1038/nature21067} {\bibfield  {journal}
  {\bibinfo  {journal} {Nature}\ }\textbf {\bibinfo {volume} {543}},\ \bibinfo
  {pages} {87} (\bibinfo {year} {2016})}\BibitemShut {NoStop}%
\bibitem [{\citenamefont {M\"uller}\ \emph {et~al.}(2007)\citenamefont
  {M\"uller}, \citenamefont {F\"olling}, \citenamefont {Widera},\ and\
  \citenamefont {Bloch}}]{Mueller2007}%
  \BibitemOpen
  \bibfield  {author} {\bibinfo {author} {\bibfnamefont {T.}~\bibnamefont
  {M\"uller}}, \bibinfo {author} {\bibfnamefont {S.}~\bibnamefont {F\"olling}},
  \bibinfo {author} {\bibfnamefont {A.}~\bibnamefont {Widera}}, \ and\ \bibinfo
  {author} {\bibfnamefont {I.}~\bibnamefont {Bloch}},\ }\href {\doibase
  10.1103/PhysRevLett.99.200405} {\bibfield  {journal} {\bibinfo  {journal}
  {Phys. Rev. Lett.}\ }\textbf {\bibinfo {volume} {99}},\ \bibinfo {pages}
  {200405} (\bibinfo {year} {2007})}\BibitemShut {NoStop}%
\bibitem [{\citenamefont {Wirth}\ \emph {et~al.}(2011)\citenamefont {Wirth},
  \citenamefont {{\"{O}}lschl{\"{a}}ger},\ and\ \citenamefont
  {Hemmerich}}]{Wirth2011}%
  \BibitemOpen
  \bibfield  {author} {\bibinfo {author} {\bibfnamefont {G.}~\bibnamefont
  {Wirth}}, \bibinfo {author} {\bibfnamefont {M.}~\bibnamefont
  {{\"{O}}lschl{\"{a}}ger}}, \ and\ \bibinfo {author} {\bibfnamefont
  {A.}~\bibnamefont {Hemmerich}},\ }\href {\doibase 10.1038/nphys1857}
  {\bibfield  {journal} {\bibinfo  {journal} {Nat. Phys.}\ }\textbf {\bibinfo
  {volume} {7}},\ \bibinfo {pages} {147} (\bibinfo {year} {2011})}\BibitemShut
  {NoStop}%
\bibitem [{\citenamefont {Isacsson}\ and\ \citenamefont
  {Girvin}(2005)}]{Isacsson2005}%
  \BibitemOpen
  \bibfield  {author} {\bibinfo {author} {\bibfnamefont {A.}~\bibnamefont
  {Isacsson}}\ and\ \bibinfo {author} {\bibfnamefont {S.~M.}\ \bibnamefont
  {Girvin}},\ }\href {\doibase 10.1103/PhysRevA.72.053604} {\bibfield
  {journal} {\bibinfo  {journal} {Phys. Rev. A}\ }\textbf {\bibinfo {volume}
  {72}},\ \bibinfo {pages} {053604} (\bibinfo {year} {2005})}\BibitemShut
  {NoStop}%
\bibitem [{\citenamefont {Liu}\ and\ \citenamefont {Wu}(2006)}]{Liu2006}%
  \BibitemOpen
  \bibfield  {author} {\bibinfo {author} {\bibfnamefont {W.~V.}\ \bibnamefont
  {Liu}}\ and\ \bibinfo {author} {\bibfnamefont {C.}~\bibnamefont {Wu}},\
  }\href {\doibase 10.1103/PhysRevA.74.013607} {\bibfield  {journal} {\bibinfo
  {journal} {Phys. Rev. A}\ }\textbf {\bibinfo {volume} {74}},\ \bibinfo
  {pages} {013607} (\bibinfo {year} {2006})}\BibitemShut {NoStop}%
\bibitem [{\citenamefont {Fischer}(2006)}]{Fischer2006}%
  \BibitemOpen
  \bibfield  {author} {\bibinfo {author} {\bibfnamefont {U.~R.}\ \bibnamefont
  {Fischer}},\ }\href {\doibase 10.1103/PhysRevA.73.031602} {\bibfield
  {journal} {\bibinfo  {journal} {Phys. Rev. A}\ }\textbf {\bibinfo {volume}
  {73}},\ \bibinfo {pages} {031602} (\bibinfo {year} {2006})}\BibitemShut
  {NoStop}%
\bibitem [{\citenamefont {Baier}\ \emph {et~al.}(2016)\citenamefont {Baier},
  \citenamefont {Mark}, \citenamefont {Petter}, \citenamefont {Aikawa},
  \citenamefont {Chomaz}, \citenamefont {Cai}, \citenamefont {Baranov},
  \citenamefont {Zoller},\ and\ \citenamefont {Ferlaino}}]{Baier2016}%
  \BibitemOpen
  \bibfield  {author} {\bibinfo {author} {\bibfnamefont {S.}~\bibnamefont
  {Baier}}, \bibinfo {author} {\bibfnamefont {M.~J.}\ \bibnamefont {Mark}},
  \bibinfo {author} {\bibfnamefont {D.}~\bibnamefont {Petter}}, \bibinfo
  {author} {\bibfnamefont {K.}~\bibnamefont {Aikawa}}, \bibinfo {author}
  {\bibfnamefont {L.}~\bibnamefont {Chomaz}}, \bibinfo {author} {\bibfnamefont
  {Z.}~\bibnamefont {Cai}}, \bibinfo {author} {\bibfnamefont {M.}~\bibnamefont
  {Baranov}}, \bibinfo {author} {\bibfnamefont {P.}~\bibnamefont {Zoller}}, \
  and\ \bibinfo {author} {\bibfnamefont {F.}~\bibnamefont {Ferlaino}},\ }\href
  {\doibase 10.1126/science.aac9812} {\bibfield  {journal} {\bibinfo  {journal}
  {Science}\ }\textbf {\bibinfo {volume} {352}},\ \bibinfo {pages} {201}
  (\bibinfo {year} {2016})}\BibitemShut {NoStop}%
\bibitem [{\citenamefont {Chomaz}\ \emph {et~al.}(2018)\citenamefont {Chomaz},
  \citenamefont {van Bijnen}, \citenamefont {Petter}, \citenamefont {Faraoni},
  \citenamefont {Baier}, \citenamefont {Becher}, \citenamefont {Mark},
  \citenamefont {W{\"{a}}chtler}, \citenamefont {Santos},\ and\ \citenamefont
  {Ferlaino}}]{Chomaz2018}%
  \BibitemOpen
  \bibfield  {author} {\bibinfo {author} {\bibfnamefont {L.}~\bibnamefont
  {Chomaz}}, \bibinfo {author} {\bibfnamefont {R.~M.~W.}\ \bibnamefont {van
  Bijnen}}, \bibinfo {author} {\bibfnamefont {D.}~\bibnamefont {Petter}},
  \bibinfo {author} {\bibfnamefont {G.}~\bibnamefont {Faraoni}}, \bibinfo
  {author} {\bibfnamefont {S.}~\bibnamefont {Baier}}, \bibinfo {author}
  {\bibfnamefont {J.~H.}\ \bibnamefont {Becher}}, \bibinfo {author}
  {\bibfnamefont {M.~J.}\ \bibnamefont {Mark}}, \bibinfo {author}
  {\bibfnamefont {F.}~\bibnamefont {W{\"{a}}chtler}}, \bibinfo {author}
  {\bibfnamefont {L.}~\bibnamefont {Santos}}, \ and\ \bibinfo {author}
  {\bibfnamefont {F.}~\bibnamefont {Ferlaino}},\ }\href {\doibase
  10.1038/s41567-018-0054-7} {\bibfield  {journal} {\bibinfo  {journal} {Nature
  Physics}\ }\textbf {\bibinfo {volume} {14}},\ \bibinfo {pages} {442}
  (\bibinfo {year} {2018})}\BibitemShut {NoStop}%
\bibitem [{\citenamefont {Petter}\ \emph {et~al.}(2018)\citenamefont {Petter},
  \citenamefont {Natale}, \citenamefont {van Bijnen}, \citenamefont
  {Patscheider}, \citenamefont {Mark}, \citenamefont {Chomaz},\ and\
  \citenamefont {Ferlaino}}]{Petter2018}%
  \BibitemOpen
  \bibfield  {author} {\bibinfo {author} {\bibfnamefont {D.}~\bibnamefont
  {Petter}}, \bibinfo {author} {\bibfnamefont {G.}~\bibnamefont {Natale}},
  \bibinfo {author} {\bibfnamefont {R.~M.~W.}\ \bibnamefont {van Bijnen}},
  \bibinfo {author} {\bibfnamefont {A.}~\bibnamefont {Patscheider}}, \bibinfo
  {author} {\bibfnamefont {M.~J.}\ \bibnamefont {Mark}}, \bibinfo {author}
  {\bibfnamefont {L.}~\bibnamefont {Chomaz}}, \ and\ \bibinfo {author}
  {\bibfnamefont {F.}~\bibnamefont {Ferlaino}},\ }\href
  {http://arxiv.org/abs/1811.12115} {} (\bibinfo {year} {2018}),\ \Eprint
  {http://arxiv.org/abs/1811.12115} {arXiv:1811.12115} \BibitemShut {NoStop}%
\bibitem [{\citenamefont {Wu}(2009)}]{Wu2009}%
  \BibitemOpen
  \bibfield  {author} {\bibinfo {author} {\bibfnamefont {C.}~\bibnamefont
  {Wu}},\ }\href {\doibase 10.1142/S0217984909017777} {\bibfield  {journal}
  {\bibinfo  {journal} {Mod. Phys. Lett. B}\ }\textbf {\bibinfo {volume}
  {23}},\ \bibinfo {pages} {1} (\bibinfo {year} {2009})}\BibitemShut {NoStop}%
\bibitem [{\citenamefont {Henkel}\ \emph {et~al.}(2010)\citenamefont {Henkel},
  \citenamefont {Nath},\ and\ \citenamefont {Pohl}}]{Henkel2010}%
  \BibitemOpen
  \bibfield  {author} {\bibinfo {author} {\bibfnamefont {N.}~\bibnamefont
  {Henkel}}, \bibinfo {author} {\bibfnamefont {R.}~\bibnamefont {Nath}}, \ and\
  \bibinfo {author} {\bibfnamefont {T.}~\bibnamefont {Pohl}},\ }\href {\doibase
  10.1103/PhysRevLett.104.195302} {\bibfield  {journal} {\bibinfo  {journal}
  {Phys. Rev. Lett.}\ }\textbf {\bibinfo {volume} {104}},\ \bibinfo {pages}
  {195302} (\bibinfo {year} {2010})}\BibitemShut {NoStop}%
\bibitem [{\citenamefont {Macr{\`{i}}}\ \emph {et~al.}(2013)\citenamefont
  {Macr{\`{i}}}, \citenamefont {Maucher}, \citenamefont {Cinti},\ and\
  \citenamefont {Pohl}}]{Macri2013}%
  \BibitemOpen
  \bibfield  {author} {\bibinfo {author} {\bibfnamefont {T.}~\bibnamefont
  {Macr{\`{i}}}}, \bibinfo {author} {\bibfnamefont {F.}~\bibnamefont
  {Maucher}}, \bibinfo {author} {\bibfnamefont {F.}~\bibnamefont {Cinti}}, \
  and\ \bibinfo {author} {\bibfnamefont {T.}~\bibnamefont {Pohl}},\ }\href
  {\doibase 10.1103/PhysRevA.87.061602} {\bibfield  {journal} {\bibinfo
  {journal} {Phys. Rev. A}\ }\textbf {\bibinfo {volume} {87}},\ \bibinfo
  {pages} {061602(R)} (\bibinfo {year} {2013})}\BibitemShut {NoStop}%
\bibitem [{\citenamefont {Kunimi}\ and\ \citenamefont
  {Kato}(2012)}]{Kunimi2012}%
  \BibitemOpen
  \bibfield  {author} {\bibinfo {author} {\bibfnamefont {M.}~\bibnamefont
  {Kunimi}}\ and\ \bibinfo {author} {\bibfnamefont {Y.}~\bibnamefont {Kato}},\
  }\href {\doibase 10.1103/PhysRevB.86.060510} {\bibfield  {journal} {\bibinfo
  {journal} {Phys. Rev. B}\ }\textbf {\bibinfo {volume} {86}},\ \bibinfo
  {pages} {060510} (\bibinfo {year} {2012})}\BibitemShut {NoStop}%
\bibitem [{\citenamefont {Georgi}(1999)}]{georgi1999lie}%
  \BibitemOpen
  \bibfield  {author} {\bibinfo {author} {\bibfnamefont {H.}~\bibnamefont
  {Georgi}},\ }\href {https://books.google.co.uk/books?id=g4yEuH5rBMUC} {\emph
  {\bibinfo {title} {Lie Algebras In Particle Physics: from Isospin To Unified
  Theories}}},\ Frontiers in Physics\ (\bibinfo  {publisher} {Avalon
  Publishing},\ \bibinfo {year} {1999})\BibitemShut {NoStop}%
\bibitem [{\citenamefont {Nemoto}\ and\ \citenamefont
  {Sanders}(2001)}]{Nemoto2002}%
  \BibitemOpen
  \bibfield  {author} {\bibinfo {author} {\bibfnamefont {K.}~\bibnamefont
  {Nemoto}}\ and\ \bibinfo {author} {\bibfnamefont {B.~C.}\ \bibnamefont
  {Sanders}},\ }\href {http://stacks.iop.org/0305-4470/34/i=10/a=309}
  {\bibfield  {journal} {\bibinfo  {journal} {J. Phys. A: Math. Gen.}\ }\textbf
  {\bibinfo {volume} {34}},\ \bibinfo {pages} {2051} (\bibinfo {year}
  {2001})}\BibitemShut {NoStop}%
\bibitem [{\citenamefont {Mermin}\ and\ \citenamefont
  {Wagner}(1966)}]{Mermin1966}%
  \BibitemOpen
  \bibfield  {author} {\bibinfo {author} {\bibfnamefont {N.~D.}\ \bibnamefont
  {Mermin}}\ and\ \bibinfo {author} {\bibfnamefont {H.}~\bibnamefont
  {Wagner}},\ }\href {\doibase 10.1103/PhysRevLett.17.1133} {\bibfield
  {journal} {\bibinfo  {journal} {Phys. Rev. Lett.}\ }\textbf {\bibinfo
  {volume} {17}},\ \bibinfo {pages} {1133} (\bibinfo {year}
  {1966})}\BibitemShut {NoStop}%
\bibitem [{\citenamefont {Hohenberg}(1967)}]{Hohenberg1967}%
  \BibitemOpen
  \bibfield  {author} {\bibinfo {author} {\bibfnamefont {P.~C.}\ \bibnamefont
  {Hohenberg}},\ }\href {\doibase 10.1103/PhysRev.158.383} {\bibfield
  {journal} {\bibinfo  {journal} {Phys. Rev.}\ }\textbf {\bibinfo {volume}
  {158}},\ \bibinfo {pages} {383} (\bibinfo {year} {1967})}\BibitemShut
  {NoStop}%
\bibitem [{\citenamefont {Gopalakrishnan}\ \emph {et~al.}(2013)\citenamefont
  {Gopalakrishnan}, \citenamefont {Martin},\ and\ \citenamefont
  {Demler}}]{Gopalakrishnan2013}%
  \BibitemOpen
  \bibfield  {author} {\bibinfo {author} {\bibfnamefont {S.}~\bibnamefont
  {Gopalakrishnan}}, \bibinfo {author} {\bibfnamefont {I.}~\bibnamefont
  {Martin}}, \ and\ \bibinfo {author} {\bibfnamefont {E.~A.}\ \bibnamefont
  {Demler}},\ }\href {\doibase 10.1103/PhysRevLett.111.185304} {\bibfield
  {journal} {\bibinfo  {journal} {Phys. Rev. Lett.}\ }\textbf {\bibinfo
  {volume} {111}},\ \bibinfo {pages} {185304} (\bibinfo {year}
  {2013})}\BibitemShut {NoStop}%
\bibitem [{\citenamefont {Watanabe}\ and\ \citenamefont
  {Brauner}(2011)}]{Watanabe2011}%
  \BibitemOpen
  \bibfield  {author} {\bibinfo {author} {\bibfnamefont {H.}~\bibnamefont
  {Watanabe}}\ and\ \bibinfo {author} {\bibfnamefont {T.}~\bibnamefont
  {Brauner}},\ }\href {\doibase 10.1103/PhysRevD.84.125013} {\bibfield
  {journal} {\bibinfo  {journal} {Phys. Rev. D}\ }\textbf {\bibinfo {volume}
  {84}},\ \bibinfo {pages} {125013} (\bibinfo {year} {2011})}\BibitemShut
  {NoStop}%
\bibitem [{\citenamefont {Watanabe}\ and\ \citenamefont
  {Murayama}(2012)}]{Watanabe2012a}%
  \BibitemOpen
  \bibfield  {author} {\bibinfo {author} {\bibfnamefont {H.}~\bibnamefont
  {Watanabe}}\ and\ \bibinfo {author} {\bibfnamefont {H.}~\bibnamefont
  {Murayama}},\ }\href {\doibase 10.1103/PhysRevLett.108.251602} {\bibfield
  {journal} {\bibinfo  {journal} {Phys. Rev. Lett.}\ }\textbf {\bibinfo
  {volume} {108}},\ \bibinfo {pages} {251602} (\bibinfo {year}
  {2012})}\BibitemShut {NoStop}%
\bibitem [{\citenamefont {Takahashi}\ and\ \citenamefont
  {Nitta}(2015)}]{Takahashi2015}%
  \BibitemOpen
  \bibfield  {author} {\bibinfo {author} {\bibfnamefont {D.~A.}\ \bibnamefont
  {Takahashi}}\ and\ \bibinfo {author} {\bibfnamefont {M.}~\bibnamefont
  {Nitta}},\ }\href {\doibase 10.1016/j.aop.2014.12.009} {\bibfield  {journal}
  {\bibinfo  {journal} {Ann. Phys.}\ }\textbf {\bibinfo {volume} {354}},\
  \bibinfo {pages} {101} (\bibinfo {year} {2015})}\BibitemShut {NoStop}%
\bibitem [{\citenamefont {Nitta}\ and\ \citenamefont
  {Takahashi}(2015)}]{Nitta2015}%
  \BibitemOpen
  \bibfield  {author} {\bibinfo {author} {\bibfnamefont {M.}~\bibnamefont
  {Nitta}}\ and\ \bibinfo {author} {\bibfnamefont {D.~A.}\ \bibnamefont
  {Takahashi}},\ }\href {\doibase 10.1103/PhysRevD.91.025018} {\bibfield
  {journal} {\bibinfo  {journal} {Phys. Rev. D}\ }\textbf {\bibinfo {volume}
  {91}},\ \bibinfo {pages} {025018} (\bibinfo {year} {2015})}\BibitemShut
  {NoStop}%
\bibitem [{Note1()}]{Note1}%
  \BibitemOpen
  \bibinfo {note} {The algorithm is as follows. Suppose we have $n$ broken
  symmetry generators $P_{1},\protect \ldots ,P_{n}$. Define the matrix:
  $\Gamma _{ij}= \protect \langle \Psi |[P_{i},P_{j}]|\Psi \protect \rangle $.
  The number of modes with even dispersion in $k$ is given by $n_{\protect
  \text {even}}=\protect \text {rank}(\Gamma )/2$ and the number of odd modes
  is $n_{\protect \text {odd}}=n-2n_{\protect \text {even}}$. The broken
  symmetries of our coherent state are $\Lambda ^{4,\protect \ldots ,8}$. We
  find that $n_{\protect \text {odd}}=1,n_{\protect \text {even}}=2$, agreeing
  with our perturbative analysis and numerical simulations.}\BibitemShut
  {Stop}%
\bibitem [{\citenamefont {Nozieres}\ and\ \citenamefont
  {Pines}(1994)}]{nozierespines2}%
  \BibitemOpen
  \bibfield  {author} {\bibinfo {author} {\bibfnamefont {P.}~\bibnamefont
  {Nozieres}}\ and\ \bibinfo {author} {\bibfnamefont {D.}~\bibnamefont
  {Pines}},\ }\href@noop {} {\emph {\bibinfo {title} {Theory of Quantum Liquids
  Volume II: Superfluid Bose Liquids.}}},\ Advanced Book Classics\ (\bibinfo
  {publisher} {Westview Press},\ \bibinfo {year} {1994})\BibitemShut {NoStop}%
\bibitem [{\citenamefont {Song}\ \emph {et~al.}(2007)\citenamefont {Song},
  \citenamefont {Semenoff},\ and\ \citenamefont {Zhou}}]{Song2007}%
  \BibitemOpen
  \bibfield  {author} {\bibinfo {author} {\bibfnamefont {J.~L.}\ \bibnamefont
  {Song}}, \bibinfo {author} {\bibfnamefont {G.~W.}\ \bibnamefont {Semenoff}},
  \ and\ \bibinfo {author} {\bibfnamefont {F.}~\bibnamefont {Zhou}},\ }\href
  {\doibase 10.1103/PhysRevLett.98.160408} {\bibfield  {journal} {\bibinfo
  {journal} {Phys. Rev. Lett.}\ }\textbf {\bibinfo {volume} {98}},\ \bibinfo
  {pages} {160408} (\bibinfo {year} {2007})}\BibitemShut {NoStop}%
\bibitem [{\citenamefont {Turner}\ \emph {et~al.}(2007)\citenamefont {Turner},
  \citenamefont {Barnett}, \citenamefont {Demler},\ and\ \citenamefont
  {Vishwanath}}]{Turner2007}%
  \BibitemOpen
  \bibfield  {author} {\bibinfo {author} {\bibfnamefont {A.~M.}\ \bibnamefont
  {Turner}}, \bibinfo {author} {\bibfnamefont {R.}~\bibnamefont {Barnett}},
  \bibinfo {author} {\bibfnamefont {E.}~\bibnamefont {Demler}}, \ and\ \bibinfo
  {author} {\bibfnamefont {A.}~\bibnamefont {Vishwanath}},\ }\href {\doibase
  10.1103/PhysRevLett.98.190404} {\bibfield  {journal} {\bibinfo  {journal}
  {Phys. Rev. Lett.}\ }\textbf {\bibinfo {volume} {98}},\ \bibinfo {pages}
  {190404} (\bibinfo {year} {2007})}\BibitemShut {NoStop}%
\bibitem [{\citenamefont {Barnett}\ \emph {et~al.}(2012)\citenamefont
  {Barnett}, \citenamefont {Powell}, \citenamefont {Gra\ss{}}, \citenamefont
  {Lewenstein},\ and\ \citenamefont {Das~Sarma}}]{Barnett2012}%
  \BibitemOpen
  \bibfield  {author} {\bibinfo {author} {\bibfnamefont {R.}~\bibnamefont
  {Barnett}}, \bibinfo {author} {\bibfnamefont {S.}~\bibnamefont {Powell}},
  \bibinfo {author} {\bibfnamefont {T.}~\bibnamefont {Gra\ss{}}}, \bibinfo
  {author} {\bibfnamefont {M.}~\bibnamefont {Lewenstein}}, \ and\ \bibinfo
  {author} {\bibfnamefont {S.}~\bibnamefont {Das~Sarma}},\ }\href {\doibase
  10.1103/PhysRevA.85.023615} {\bibfield  {journal} {\bibinfo  {journal} {Phys.
  Rev. A}\ }\textbf {\bibinfo {volume} {85}},\ \bibinfo {pages} {023615}
  (\bibinfo {year} {2012})}\BibitemShut {NoStop}%
\bibitem [{\citenamefont {Nelson}\ and\ \citenamefont
  {Pelcovits}(1977)}]{NelsonPelcovits1977}%
  \BibitemOpen
  \bibfield  {author} {\bibinfo {author} {\bibfnamefont {D.~R.}\ \bibnamefont
  {Nelson}}\ and\ \bibinfo {author} {\bibfnamefont {R.~A.}\ \bibnamefont
  {Pelcovits}},\ }\href {\doibase 10.1103/PhysRevB.16.2191} {\bibfield
  {journal} {\bibinfo  {journal} {Phys. Rev. B}\ }\textbf {\bibinfo {volume}
  {16}},\ \bibinfo {pages} {2191} (\bibinfo {year} {1977})}\BibitemShut
  {NoStop}%
\bibitem [{\citenamefont {Fellows}\ \emph {et~al.}(2012)\citenamefont
  {Fellows}, \citenamefont {Carr}, \citenamefont {Hooley},\ and\ \citenamefont
  {Schmalian}}]{Fellows2012}%
  \BibitemOpen
  \bibfield  {author} {\bibinfo {author} {\bibfnamefont {J.~M.}\ \bibnamefont
  {Fellows}}, \bibinfo {author} {\bibfnamefont {S.~T.}\ \bibnamefont {Carr}},
  \bibinfo {author} {\bibfnamefont {C.~A.}\ \bibnamefont {Hooley}}, \ and\
  \bibinfo {author} {\bibfnamefont {J.}~\bibnamefont {Schmalian}},\ }\href
  {\doibase 10.1103/PhysRevLett.109.155703} {\bibfield  {journal} {\bibinfo
  {journal} {Phys. Rev. Lett.}\ }\textbf {\bibinfo {volume} {109}},\ \bibinfo
  {pages} {155703} (\bibinfo {year} {2012})}\BibitemShut {NoStop}%
\bibitem [{Note2()}]{Note2}%
  \BibitemOpen
  \bibinfo {note} {In preparation.}\BibitemShut {Stop}%
\bibitem [{\citenamefont {Chiacchio}\ and\ \citenamefont
  {Nunnenkamp}(2018)}]{RodriguezChiacchio2018}%
  \BibitemOpen
  \bibfield  {author} {\bibinfo {author} {\bibfnamefont {E.~I.~R.}\
  \bibnamefont {Chiacchio}}\ and\ \bibinfo {author} {\bibfnamefont
  {A.}~\bibnamefont {Nunnenkamp}},\ }\href {\doibase
  10.1103/PhysRevA.98.023617} {\bibfield  {journal} {\bibinfo  {journal} {Phys.
  Rev. A}\ }\textbf {\bibinfo {volume} {98}},\ \bibinfo {pages} {023617}
  (\bibinfo {year} {2018})}\BibitemShut {NoStop}%
\end{thebibliography}%
\bibliographystyle{apsrev4-1}

\end{document}